\documentclass[final,3p,times,authoryear]{elsarticle}
\usepackage{textcomp}
\usepackage{amssymb}
\usepackage[mathlines]{lineno}
\usepackage[latin1]{inputenc}   
\usepackage{amsmath}            
\usepackage{epstopdf}          
\usepackage{flushend}   
\usepackage{geometry}
\usepackage{blindtext}
\usepackage[noend]{algpseudocode}
\usepackage{threeparttable}
\usepackage{booktabs}
\usepackage{lineno}
\usepackage[labelfont=it,
            textfont=it,
            font=normalsize]{caption}
\usepackage{color}
\usepackage{natbib}
\usepackage[figuresright]{rotating}
\usepackage{subfigure}
\usepackage{enumitem}[align=left]
\usepackage[section]{placeins}
\usepackage{float}
\usepackage{footnote}
\usepackage{multirow}
\usepackage[table,xcdraw]{xcolor}
\usepackage{diagbox}
\usepackage{multicol}
\usepackage[colorlinks]{hyperref}
\makeatletter
\def\UrlAlphabet{%
      \do\a\do\b\do\c\do\d\do\e\do\f\do\g\do\h\do\i\do\j%
      \do\k\do\l\do\m\do\n\do\o\do\p\do\q\do\r\do\s\do\t%
      \do\u\do\v\do\w\do\x\do\y\do\z}
\g@addto@macro{\UrlBreaks}{\UrlOrds}
\g@addto@macro{\UrlBreaks}{\UrlAlphabet}
\makeatother

\makeatletter
\newcommand{\algorithmfootnote}[2][\footnotesize]{%
  \let\old@algocf@finish\@algocf@finish
  \def\@algocf@finish{\old@algocf@finish
    \leavevmode\rlap{\begin{minipage}{\linewidth}
    #1#2
    \end{minipage}}%
  }%
}
\makeatother

\def\mathbi#1{\textbf{\em #1}}

\usepackage{amsmath,amssymb,amsfonts,bm,mathtools}
\usepackage[ruled,linesnumbered]{algorithm2e}
\setcitestyle{authoryear,round}
\journal{Transportation Research Part E: Logistics and Transportation Review}

\begin{document}

\begin{frontmatter}


\title{A Route Network Planning Method for Urban Air Delivery}

\author[First]{Xinyu He}
\ead{xinyuhe5-c@my.cityu.edu.hk}

\author[First,Second]{Fang He}
\ead{fanghe7-c@my.cityu.edu.hk}

\author[First,Third]{Lishuai Li\corref{cor1}}
\ead{lishuai.li@cityu.edu.hk}

\author[Fourth]{Lei Zhang}
\ead{zhanglei@antwork.link}

\author[Second]{Gang Xiao}
\ead{xiaogang@sjtu.edu.cn}

\cortext[cor1]{Corresponding author}

\address[First]{School of Data Science, City University of Hong Kong, Hong Kong Special Administrative Region}
\address[Second]{School of Aeronautics and Astronautics, Shanghai Jiao Tong University, Shanghai, China}
\address[Third]{Air Transport and Operations, Faculty of Aerospace Engineering, Delft University of Technology, Delft, Netherlands}
\address[Fourth]{Antwork Technology Co., Ltd, Hangzhou, China}

\begin{abstract}
High-tech giants and start-ups are investing in drone technologies to provide urban air delivery service, which is expected to solve the last-mile problem and mitigate road traffic congestion. However, air delivery service will not scale up without proper traffic management for drones in dense urban environment. Currently, a range of Concepts of Operations (ConOps) for unmanned aircraft system traffic management (UTM) are being proposed and evaluated by researchers, operators, and regulators. Among these, the tube-based (or corridor-based) ConOps has emerged in operations in some regions of the world for drone deliveries and is expected to continue serving certain scenarios that with dense and complex airspace and requires centralized control in the future. Towards the tube-based ConOps, we develop a route network planning method to design routes (tubes) in a complex urban environment in this paper. In this method, we propose a priority structure to decouple the network planning problem, which is NP-hard, into single-path planning problems. We also introduce a novel space cost function to enable the design of dense and aligned routes in a network. The proposed method is tested on various scenarios and compared with other state-of-the-art methods. Results show that our method can generate near-optimal route networks with significant computational time-savings.
\end{abstract}

\begin{keyword}
urban air delivery\sep urban air mobility\sep unmanned aircraft system traffic management\sep multi-path planning

\end{keyword}

\end{frontmatter}

\section{Introduction}

Urban parcel delivery via drones belongs to the broad concept of Urban Air Mobility (UAM). It is a rapidly emerging field in research and business development, with prospects to ease urban traffic congestion, connect remote areas with great agility, lower labor costs in logistics, and ensure goods delivery in emergencies \citep{Duvall2019,chung2021applications,Skrinjar2019,rajendran2020air,kellermann2020drones, kitonsa2018significance,lemardele2021potentialities}. Although the number of operations of drone deliveries is not large yet, the global market of drones and electric aircraft operations is expected to increase to tens of billions of USD in the early 2030s estimated by McKinsey \citep{Kersten2021}, and around 1 trillion USD by 2040 estimated by Morgan Stanley \citep{morgan2021}. To scale up the operations, a key challenge lies in how to manage a large volume of drone operations in a dense urban environment efficiently and safely. Therefore, there are several ongoing R\&D programs to explore concepts of operation, data exchange requirements, and a supporting framework to enable drone operations at low altitudes in airspace where traditional air traffic services are not provided, such as NASA/FAA unmanned aircraft system traffic management (UTM) \citep{nasa2021}, SESAR U-space \citep{SESARJoint2019}, Singapore uTM-UAS \citep{mohamed2018preliminary}, J-UTM \citep{ushijima2017utm}, etc. 

A range of UTM Concepts of Operations (ConOps) for traffic and airspace management are being explored and studied \citep{SESARJoint2019,Bauranov2021}. Each ConOps has its own advantages and disadvantages. As stated in \citep{EUROCONTROL2018}, different ConOps may co-exist in the future and each suits certain scenarios. For example, free-flight-based operations spread the traffic over the whole airspace to reduce the number of potential conflicts \citep{Hoekstra2002,Jardin2005,Krozel2001,Yang2018}. It allows each drone to follow its optimal path, detect and avoid other flights. Structure-based operations use traffic flow constraints to reduce airspace complexity and management workload \citep{Krozel2001,OctavianThor2009,Sunil2015}. The structures may include layers, zones, and tubes (or air corridors); they separate drones and organize traffic flows to reduce potential conflicts \citep{BinMohammedSalleh2018,Jang2017,Kopardekar2014,Kopardekar2016}.  

\begin{figure}[ht]
\centering
  \includegraphics[width=15cm]{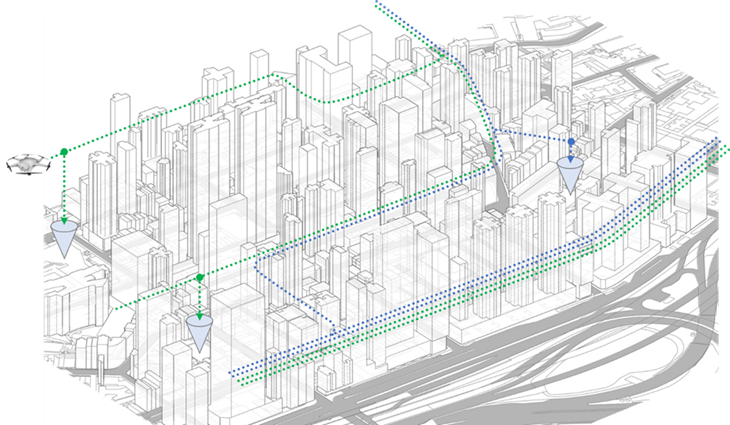}\\
  \caption{An imaginary scenario of urban air delivery route network}\label{fig1}
\end{figure}

Among these ConOps, the concept of tube-based operations has been proposed by Eurocontrol \citep{EUROCONTROL2018} and NASA \citep{Jang2017} as a kind of structure-based operations. Tubes are pre-planned routes "to cover for higher traffic demands, specifically in areas where high volume flight routes occur or there are needs to manage routing to cater for safety, security, noise, and privacy issues" \citep{EUROCONTROL2018}. They are also referred to as "structured routes", or "air corridors." These structured routes can follow rivers, railway lines, or other geographical areas where there is minimal impact on people on the ground. An imaginary tube-based route network scenario is shown in Figure \ref{fig1}. Recently, tube-based operations have been implemented for drone deliveries in a few cities in China. For example, a start-up company, Antwork Technology, has been operating drone parcel deliveries on the tube-based inner-city network in Hangzhou, China, since 2019 when it obtained the world's first approval and business license from the Civil Aviation Administration of China (CAAC) for operating commercial drone deliveries in urban areas. We expect the tube-based operations will continue to grow for regions with dense and complex airspace and that requires centralized control in the future. 

To support the tube-based operations, we develop a route network planning method to design routes (tubes) in a complex urban environment in this paper. In general, the design of a route network for drone deliveries includes 1) strategic design, 2) tactical design. The strategic design is mainly driven by business considerations, e.g. identifying areas with high demand, needing medical / emergency responses, or lacking other transportation infrastructure, to select the vertiport locations and the network types to best serve a market. This kind of network design is outside the scope of this work, which is expected to be decided before planning the routes as presented in this work. The tactical design focuses on path planning for drone operations considering operational constraints, such as air traffic management restrictions, safety requirements, public acceptance, noise ordinances, etc. The tactical design of a route network for drone operations in an urban environment is the focus of this work. 

There are many existing path planning methods in the literature. However, to the best of our knowledge, none of them can be directly used to design a route network for urban air delivery at a city level due to the computational complexity involved in real-world scenarios. The problem of designing an optimal network with spatially separated routes is NP-hard \citep{Yu2013}. Path finding for one route is coupled with path finding for other routes. One route is planned with the shortest length (or the lowest cost), but other routes may need to take a significant detour to avoid any conflict, resulting in a sub-optimal solution at the system level. The computational complexity increases exponentially as the number of conflicts among routes increases. In addition, there is an unique requirement on the design of a drone delivery network in an urban area, which is the spread of a route network needs to be minimized to reduce the impact of high volume drone operations on safety, noise, and privacy, etc. Therefore, the routes in a network need to be aligned and close to each other as much as possible, which would reduce the total size of areas been affected by drone operations and increase the utilization of limited airspace in urban areas. The alignment of traffic flows would also decrease the traffic complexity in airspace and reduce the probability of separation loss \citep{Hoekstra2018}. However, no existing studies on path planning have considered network spread and airspace utilization.

To tackle these challenges, we propose a sequential path finding method to design a tube-based route network for drone operations in an urban environment considering airspace utilization. In this method, we propose a prioritization structure to decouple the network planning problem into a set of single-path planning problems to generate paths sequentially. We set the prioritization rules to rank origin-destination (OD) pairs to ensure important routes with higher priority in planning for better system performance. To obtain better airspace utilization, we introduce a space cost function in the underlying path-finding algorithm to build a network with dense and aligned routes. The contributions of this work are three folds:
\begin{itemize}
\setlength{\itemsep}{0pt}
\setlength{\parsep}{0pt}
\setlength{\parskip}{0pt}
\item[1)]
A sequential route network planning method with prioritization is developed to support tube-based operations for Unmanned Aerial Vehicles (UAVs) traffic management. The proposed prioritization framework can solve the NP-hard problem by decoupling the multi path-finding problem into sequential single path-finding problems to generate results fast.
\item[2)]
A new technique, referred to as space cost function, is developed to enable the design of dense route networks with improved airspace utilization.
\item[3)]
Comparative experiments and real-world applications are carried out to benchmark with other state-of-the-art methods and show the performance of the proposed method under various settings in real-world scenarios.
\end{itemize}

The remainder of this paper is structured as follows. An overview of related academic literature is provided in Section 2. The statement of the problem is illustrated in Section 3. The proposed method for route network planning is explained in Section 4. We evaluate the proposed method in testing scenarios and a real-world scenario in Section 5. Finally, Section 6 concludes the research findings and proposes future work.

\section{Literature Review}
On drone delivery problem, a large number of studies focuses on vehicle routing problem with drones \citep{shen2020synergistic} and traffic management problem for drones. The vehicle routing problems involve how to coordinate delivery trucks and drones to deliver small parcels to geographically distributed customers \citep{ moshref2020design,Ha2018,Karak2019, Murray2015, Murray2020, Sacramento2019,Schermer2019,zhang2021humanitarian}. The traffic management problem is about how drones fly safely to finish tasks. It includes three steps. One is to plan a trajectory for each drone operation \citep{yang20083d,cekmez2014uav,sonmez2015optimal,zhao2018survey, wu2021swarm}, one is to detect conflicts when drones follow these trajectories \citep{tan2019evolutionary,islami2017large,kuchar2000review}, one is to resolve conflicts among drones if conflicts appear \citep{yang2020scalable,bertram2021scalable,yang2021autonomous,tang2021automated}. 

In this paper, we focus on the problem of designing a network with multiple routes in an urban environment; each route connects an origin and a destination and is spatially separated from each other. Related studies can be broadly grouped in to two categories : the single-path finding (SPF) problem and the multi-path finding (MPF) problem.

\subsection{Single-Path Finding}
SPF involves moving a single agent to a destination among a set of obstacles. SPF is a well-studied area and many algorithms are proposed to find a path for the single agent. 

Graph-search-based algorithms require a discretization of the state space, they search over the graph to find a path. The Dijkstra's algorithm was the pioneer \citep{Dijkstra1959, liu2020integrating}, it searches the minimum cost paths between two points in the graph by comparing all possible combinations. A* is an advancement of the Dijkstra's algorithm \citep{Hart1968}, it uses a heuristic function to estimate the cost of reaching the goal to reduce computational times. Many variants of A* were developed, like Dynamic A* (D*) \citep{Koenig2005}, Theta* \citep{Daniel2010}. For these methods, some solutions might be missed if the resolution of the discretization is large, and these methods do not guarantee to find an optimal path because of the heuristic nature, unless the heuristic is admissible, i.e., it never overestimates the cost of reaching the goal. In summary, these methods are efficient to find near-optimal paths with an appropriate resolution of the discretization in terms of the large size of the designing space.

Sampling-based algorithms do not require a discretization of the state space, they generate some possible paths by randomly adding points in the environment into a tree until some solution is found or time expires. Two prominent examples are Probabilistic Roadmap Method (PRM) \citep{Kavraki1996} and Rapidly-Exploring Random Tree (RRT) \citep{Lavalle1998}. The PRM method first samples points in the environment and connect points to form a roadmap, and then it searches for a feasible path using this constructed roadmap. The RRT method grows a unidirectional search tree from the starting point until a tree branch hits the goal point. The RRT method guarantees finding a feasible solution as the number of iterations approaches infinity, however, the paths are often much longer than the optimal paths. Though its variant RRT* \citep{karaman2011sampling} is asymptotically optimal, the convergence rate is very slow. A comparison shows \citep{zammit2018comparison} that A*'s path length is more optimal and generation time is shorter than RRT for path planning for UAVs. In summary, sampling-based methods are still efficient to find a feasible path in terms of the large size of the designing space, but the path is extremely sub-optimal. 

There are also some methods like mathematical optimization-based algorithms, neural network-based algorithms, nature-inspired algorithms. The mathematical optimization-based algorithms formulate the path finding problem as binary linear programming problems \citep{Chamseddine2012} or mixed-integer programming problems \citep{Culligan2007}, and use high-quality solvers to solve these programming problems. Neural network-based algorithms \citep{yang2000efficient, dezfoulian2013generalized, singh2019neural} use neural networks to model complex environments. Natural inspired algorithms, like genetic algorithms \citep{hu2004knowledge}, particle swarm optimization \citep{masehian2010multi}, ant colony optimization \citep{englot2011multi} are also successfully applied for path finding. However, these methods are time consuming, they are not efficient to find a path in terms of the large size of the designing space.

In summary, if the size of the designing space is large, A* and its variants are the best choices considering computational time and the optimality of the path. 

\subsection{Multi-Path Finding}
MPF involves navigating the agents to their respective targets while avoiding collisions with obstacles and other agents. Such problems are NP-hard \citep{Yu2013}. There are two different Multi-path finding problems based on the different cost function to minimize, one is sum-of-cost and another is makespan \citep{Felner2017}. For sum-of-cost problems, the sum of the total time steps or total distance for all agents should be minimized. For makespan problems, the maximum of the individual path costs should be minimized. Designing tube-based route networks is similar to the sum-of-cost problems, but there is a major difference that the sum-of-cost problems only require paths to be temporal conflict-free, but tube-based route networks further require paths to be spatial conflict-free, i.e., no path can appear at the same place even at different times.

Traditional MPF algorithms fall into two settings: centralized and distributed. In a distributed setting, each agent has its computing power and it does not have full knowledge of other agents. Some work has been done for the distributed setting \citep{Bhattacharya2010,Gilboa2006,Grady2010}. By contrast, the centralized setting assumes a single central computing power which needs to find a solution for all agents. The scope of this paper is limited to the centralized setting, thus here we provide a review on MPF algorithms in a centralized setting grouped into two categories: optimal solvers and sub-optimal solvers.

\subsubsection{Optimal Solvers for Centralized MPF}
The optimal solvers can generate optimal paths and they are complete in theory. However, the dimension of the problem explode exponentially as the number of agents increases. Thus, they are often used to generate routes for a small number of agents. These solvers can be broadly classified into A*-based, reduction-based, and two-level-based methods.

A* can be well suited to solve MPF sum-of-cost problems by taking k-agent state space. Each state includes k agents as a vector, e.g., in the start state each agent is located at its start point. For each state, there are $4^k$ neighbors to explore if each agent can move in four cardinal directions. Even for 10 agents, there are $4^{10}\approx 10^6$ neighbors for the start state. As a result, the original A*-based method is computationally infeasible to solve real-world problems. A few techniques are developed to speed up A*. For example, independence detection divides the agents into independent groups and solves these groups separately \citep{Standley2010, Wagner2015}, here two groups are independent if the optimal solution for each group does not conflict with the other. Another technique is related to surplus nodes, which are nodes generated but never be expanded to find an optimal solution. Avoiding generating the surplus nodes makes a substantial speedup \citep{Standley2010, Felner2012, Goldenberg2014}. In summary, though these techniques provide exponential speedup for A*-based methods, solution quality still degrades rapidly and computational time increases fast as the agent density increase.

Reduction-based methods reduce the MPF problem to standard known problems. Examples include network flow problems with integer linear programming formulations \citep{Yu2016,Yu2013b}, Boolean satisfiability \citep{Surynek2012}, and constraint satisfaction problems \citep{Ryan2010}. These methods are designed for the MPF makespan problems, and they are inefficient or even inapplicable for MPF sum-of-cost problems. 

Two-level-based methods introduce a tree structure to solve MPF sum-of-cost problems. Each node in the tree includes a solution for all routes, At the high level, these methods search over the tree to find the optimal solution, and then the low-level search is invoked to generate a node of the tree. Typical two-level-based methods include increasing cost tree search \citep{Sharon2013} and conflict-based search (CBS) \citep{Sharon2015}. CBS is a state-of-the-art optimal solver. It uses a constraint tree (CT) for search. In a CT, each node includes a set of constraints on agents, a solution for all agents that satisfies the constraints, and the cost of the solution. The root node of the CT contains no constraint, and the paths in the solution are individually optimal if ignoring other agents. The tree is built progressively from root nodes to leaf nodes. In each step, the high-level search assigns a conflict as a constraint for an agent to satisfy. On the leaf nodes, all conflicts are solved, so the solutions are all feasible paths. The leaf node with the minimum cost has optimal paths. In summary, these methods are efficient and find optimal paths for small problems. If there are many agents and many conflicts to solve, their computational times also increase very fast. 

\subsubsection{Sub-Optimal Solvers for Centralized MPF}
The sub-optimal solvers are commonly used to generate feasible routes quickly for a large number of agents for sum-of-cost problems. By decomposing the problem into several sub-problems, the computational time can be significantly reduced. But most of the time they can only find sub-optimal paths and, in some cases, completeness is sacrificed to improve time complexity. The sub-optimal solvers can be roughly classified as rule-based methods, search-based methods, and hybrid methods.

Rule-based solvers make specific agent-movement rules for different scenarios and they do not need massive search steps. They usually require special properties of the graphs to guarantee completeness. The algorithm by \citep{kornhauser1984coordinating} guarantees the completeness for all different graphs but the implementation is complex. BIBOX \citep{Surynek2009} is complete for bi-connected graphs. Push-and-Swap \citep{Luna2011} uses "swap" macro to swap locations between two adjacent dependent agents. Its variants Push-and-Rotate \citep{DeWilde2014}, Push-and-Spin \citep{Alotaibi2018}, etc., use more macros to handle complex situations and graphs. However, deadlocks often happen in narrow corridors and inter-agent collision-free may not be guaranteed, so the algorithms may fail to find paths even the paths exist (incompleteness). In summary, rule-based methods are efficient if there are a large number of agents, but the generated results are often far away from optimal.

Search-based methods search over the graph to find feasible solutions. The solutions are often near-optimal and sometimes but they are not complete for many cases. Prioritized approaches are a kind of prominent search-based methods. They plan routes sequentially and treat positions of previously planned agents as moving obstacles to avoid any collision \citep{VanDenBerg2005}. Hierarchical Cooperative A* (HCA*) \citep{Silver2005} is a typical prioritized approach. HCA* plans one route for one agent at a time according to a predefined order, and it stores each route into a reservation table after the route is planned. The previous paths, i.e., entries in the reservation table, are impassable for later agents. Windowed HCA* (WHCA*) \citep{Silver2005} runs in a similar way but it uses plan-move cycles to dynamically generate routes. In each planning phase, each agent reserves the next W steps by the order; in each moving phase, each agent follows the reserved paths by K ($K\leq W$) steps, then the next cycle starts in the current point. Conflict-oriented WHCA* (CO-WHCA*) \citep{bnaya2014conflict} places the window, i.e., the reservation for the next W steps, only around conflicts. The choice of priorities has a great impact on the performance of planned routes \citep{Warren1990} and there exist different strategies. Arbitrary order is applied in HCA* and the planning phase in each cycle in WHCA*. The decreasing order of the shortest path length for each agent is taken as the criterion in \citep{VanDenBerg2005}. A winner-determination strategy is taken in CO-WHCA*, where all possible orders are estimated for every conflict and the best one is selected. Several search-based methods are bounded sub-optimal solvers \citep{barer2014suboptimal, cohen2019optimal, cohen2016improved}. Most of them are variants of conflict based search, they provide bounded sub-optimal solutions by relaxing the assumptions and conditions. In summary, prioritized approaches provide near-optimal solutions and they are still efficient when there are many agents. Bounded sub-optimal methods improve computational time compared to optimal methods, but they are still not efficient for a large number of agents.

Hybrid methods take both specific agent-movement rules and massive search steps. In the first phase, a path is planned for each agent individually with obstacle avoidance by using SPF algorithms while other agents' paths are ignored at this phase. In the next phase, the agents coordinate to ensure no inter-agent collision occur along the paths. Common coordination methods include modification of geometric paths, modification of velocity profiles, and delay in the startup time of agents \citep{Kant1986, Leroy1999, Li2005, ODonnell1989, Peng2005,Saha2006,Sanchez2002}. These coordination schemes are the rules for solving conflicts. In summary, these methods can find paths fast if there are many agents, but most of them use time dimension to avoid collision and they cannot guarantee paths are spatial conflict-free.

In summary, none of the existing multi-path finding methods can readily solve the route network planning problem in this paper. The problem is NP-hard. The search space for the drone network design is large considering the size of the design space, cluttered obstacles, complexity of risk levels, and the coupled complexity of many routes to be planned. The optimal solvers are inefficient to solve the problem as it suffers the curse of dimensionality. Rule-based sub-optimal methods and hybrid sub-optimal methods are much efficient, but their solutions are often too far away from optimal. Also, they require to use the time dimension to solve conflicts. Spatial intersections may still exist in the generated network. Prioritized approaches are applicable and efficient, and their results are near-optimal. Therefore, following the prioritized approaches, we develop a set of prioritization rules and integrate them into a sequential route planning algorithm for drone route network planning in an urban environment in this paper.

\section{Problem Statement}
\begin{figure}[ht]
\centering
  \includegraphics[width=15cm]{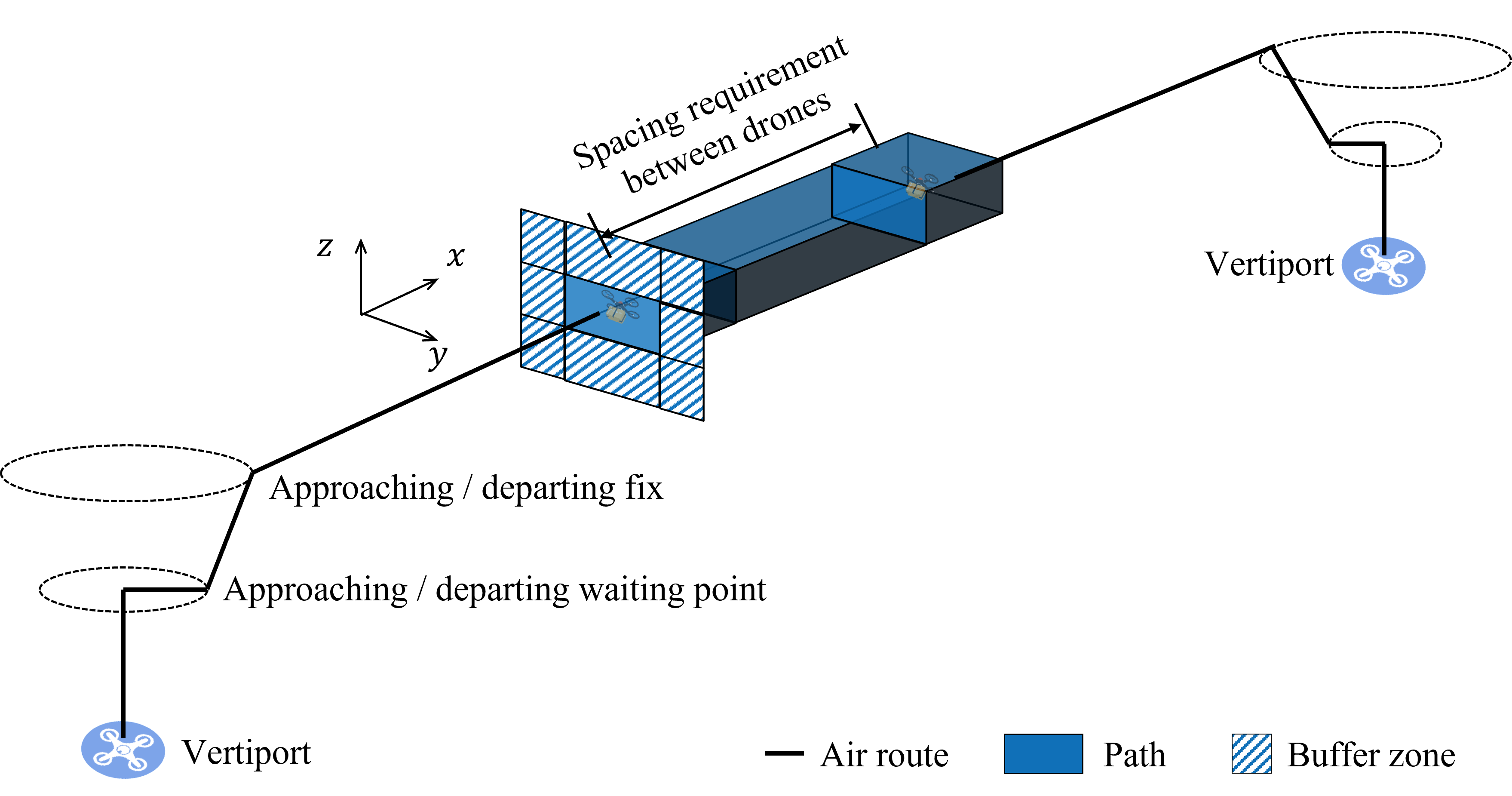}\\
  \caption{Components of a route network}\label{fig2}
\end{figure}

\begin{figure}[ht]
\centering
  \includegraphics[width=15cm]{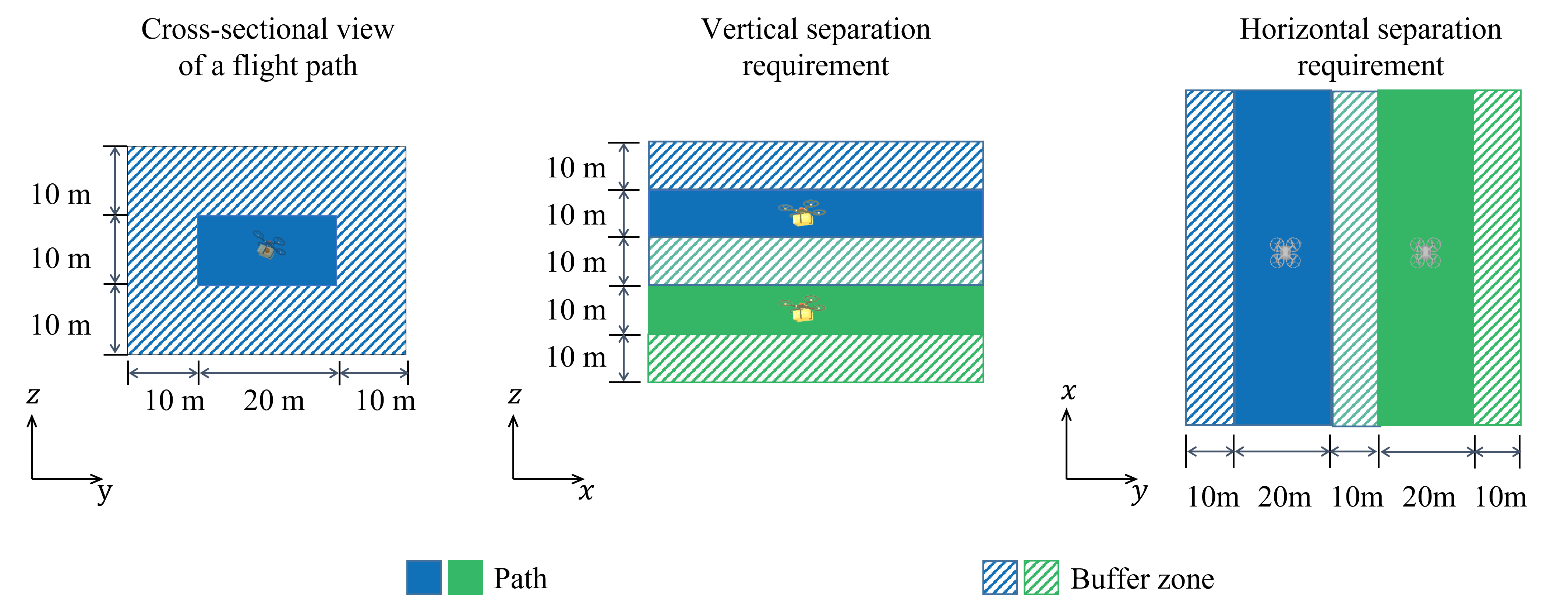}\\
  \caption{Dimension of a flight path and its separation requirement}\label{fig3}
\end{figure}

\begin{figure}[ht]
\centering
  \includegraphics[width=6cm]{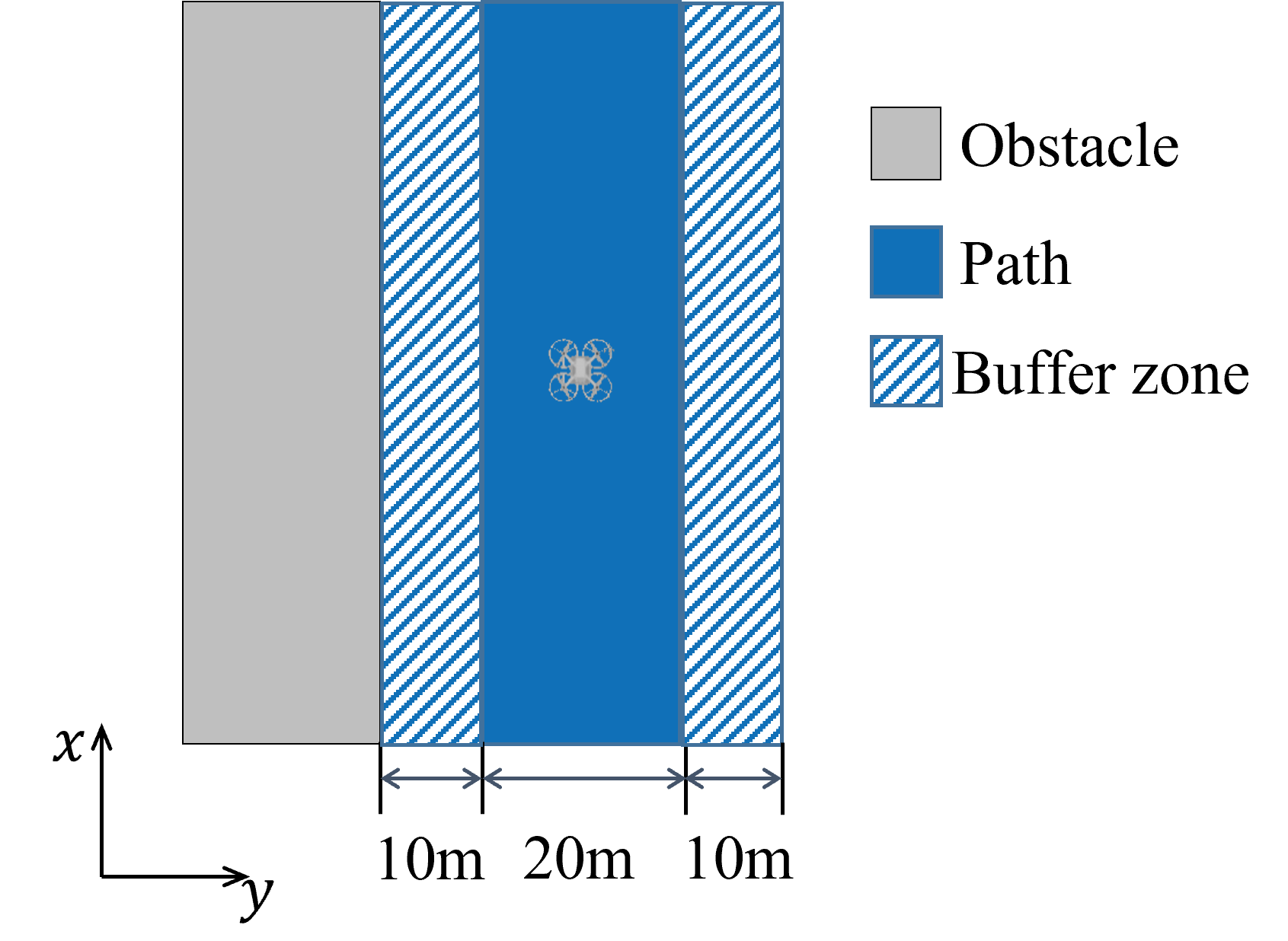}\\
  \caption{Separation requirement between a path and an obstacle}\label{fig4}
\end{figure}

This section defines the drone delivery network that we aim to design. An illustration of the network is shown in Figure \ref{fig2}. Air routes are unidirectional paths established among approaching and departing fixes over Origin-Destination vertiports. Drones can fly sequentially in a path following the minimum spacing requirement. The width and height of a path are 20 meters and 10 meters, which is determined based on drone position uncertainties and measurement errors. There is a layer of airspace surrounding the path that serves as "buffer zones" with a width of 10 m, as shown in Figure \ref{fig3}. No other path or obstacle is allowed in the buffer zone, and the vertical and horizontal separation requirements are shown in Figure \ref{fig3} and Figure \ref{fig4}. However, the buffer zones of different paths can be overlapped as shown in Figure \ref{fig3}. 

For the design of a route network, individual paths are expected to expose minimum risks to the ground and to impose minimum energy consumption of drone operations. On a network level, the spread of the route network should be minimized. Besides, the computational time for generating a route network should also be short.

\section{Methodology for Route Network Planning}
\subsection{Overview}
To design a route network for an urban area as described in Section 3, we develop a novel sequential route network planning method considering airspace utilization. In this method, we propose a prioritization structure to decouple the network planning problem into a set of single-path planning problems to generate paths sequentially. We set the prioritization rules to rank origin-destination (OD) pairs to ensure important routes with higher priority in planning for better system performance. To obtain better airspace utilization, we introduce a space cost function in the underlying path-finding algorithm to build a network with dense and aligned routes.

\begin{figure}[ht]
\centering
  \includegraphics[width=9cm]{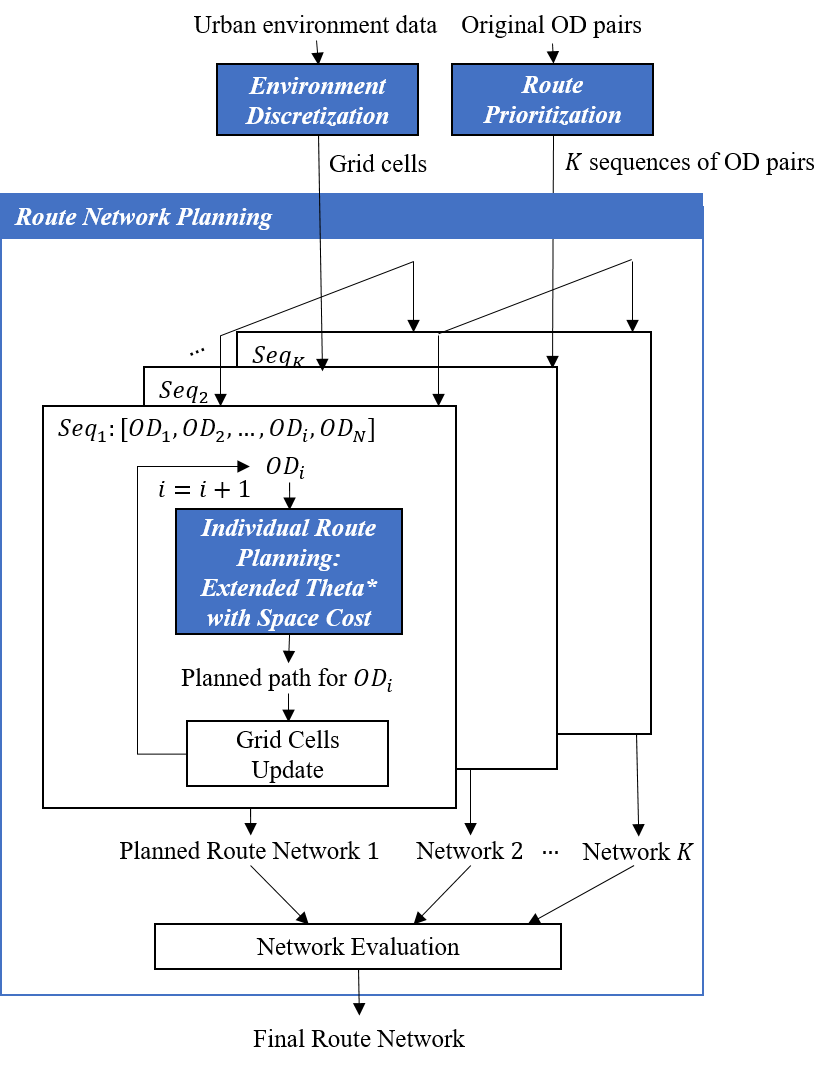}\\
  \caption{Overall workflow of the proposed route network planning method}
\label{fig5}
\end{figure}

The proposed route network planning method is composed of four modules: Environment Discretization, Individual Route Planning, Route Prioritization, and Network Planning. The \textit{Environment Discretization} module generates graphs for searching, the \textit{Route Prioritization} module generates multiple ordered route sequences, and the \textit{Network Planning} module generates route networks based on the graphs and the route sequences. In generating a route network, the \textit{Individual Route Planning} module is iteratively conducted to generate a route network for each route sequence. The \textit{Network Evaluation} module selects the route network that has a minimum cost from all generated route networks, then it checks the risk for each path in the selected route network. If all paths satisfy the risk requirement, the selected route network will be returned as the final route network; otherwise, the method fails to find a feasible route network. The overall workflow is illustrated in Figure \ref{fig5}. The associated algorithm, named as \textit{Sequential Extended Theta* with Route Prioritization and Space Cost}, is shown in Algorithm \ref{alg1}.

\FloatBarrier

\begin{algorithm}[H]
    \renewcommand{\thealgocf}{1}
    \SetAlgoLined
    \DontPrintSemicolon
    \SetKwInOut{Input}{Input}
    \SetKwInOut{Output}{Output}
    \SetKwProg{Fn}{}{}{}
    \Input{OD pairs $\{OD_{i}\}_{i=1}^{N}$, environment $E$}
    \Output{A route network $\{R_{i}\}_{i=1}^{N}$}
    /* Environment Discretization */\;
    Perform space discretization on $E$;\;
    Generate grid cells to compose graph $G(V,E)$;\;
    /* Route Prioritization */\;
    Generate $K$ OD pair sequences $\{Seq_k\}_{k=1}^K$from original OD pairs $\{OD_{i}\}_{i=1}^{N}$;\;
    /* Network Planning */\;
    \For{$Seq_k$ in $\{Seq_k\}_{k=1}^K$}{
    \For{$OD_i$ \textnormal{in} $Seq_k$}{
    /*Individual planning*/\;
    Run Extended Theta* with Space Cost on $OD_i$ to get planned paths $R_{i}^{k}$;\;
    /*Update grid cells*/\;
    Set path cells for $R_{i}^{k}$ as unreachable (impassable for obstacles and $R_{j>i}^{k}$);\;
    Set buffer zone cells for $R_{i}^{k}$ as reserved (impassable for obstacles and paths of $R_{j>i}^{k}$, passable for buffer zones of $R_{j>i}^{k}$);
        }
    Store the route network $\{R_{i}^{k}\}_{i=1}^{N}$;
    }
    /* Network Evaluation */\;
    Find $\{R_{i}\}_{i=1}^{N}$ with lowest costs from $K$ route networks $\{\{R_{i}^{k}\}_{i=1}^{N}\}_{k=1}^K$\;
    Check risk for each path;\;
    \uIf{a subset of the paths do not satisfy the risk requirement}{
        return "fail to generate a feasible route network"
    }\Else{
        Return the route network  $\{R_{i}\}_{i=1}^{N}$
    }
    
    \caption{Route Network Planning: Sequential Extended Theta* with Route Priority and Space Cost}
    \label{alg1}
\end{algorithm}

\subsection{Environment Discretization}

\begin{figure}[ht]
\centering
  \includegraphics[width=12cm]{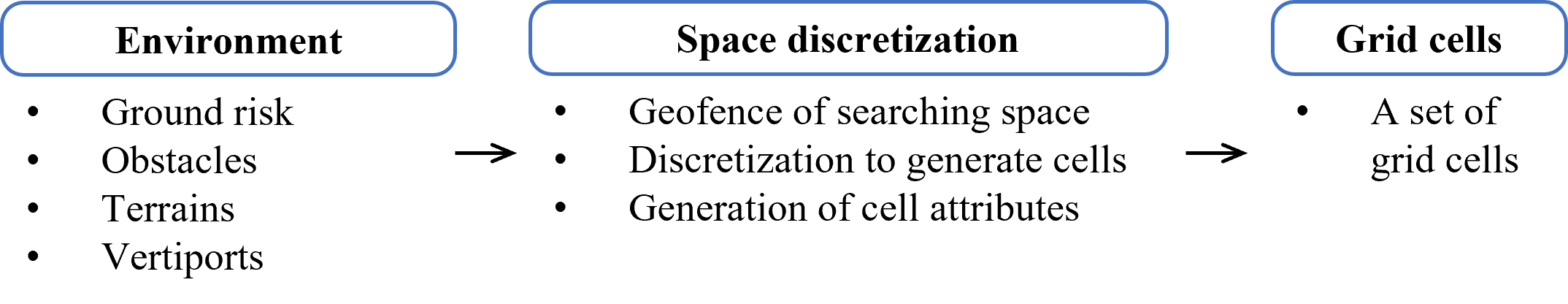}\\
  \caption{Flowchart of environment discretization}
\label{fig6}
\end{figure}

This module aims to generate a grid graph for the network planning. It discretizes the environment into 3D cubic grid cells to compose a grid graph $G(V,E)$. The process is shown in Figure \ref{fig6}.

\FloatBarrier

\begin{figure}[ht]
\centering
  \includegraphics[width=8cm]{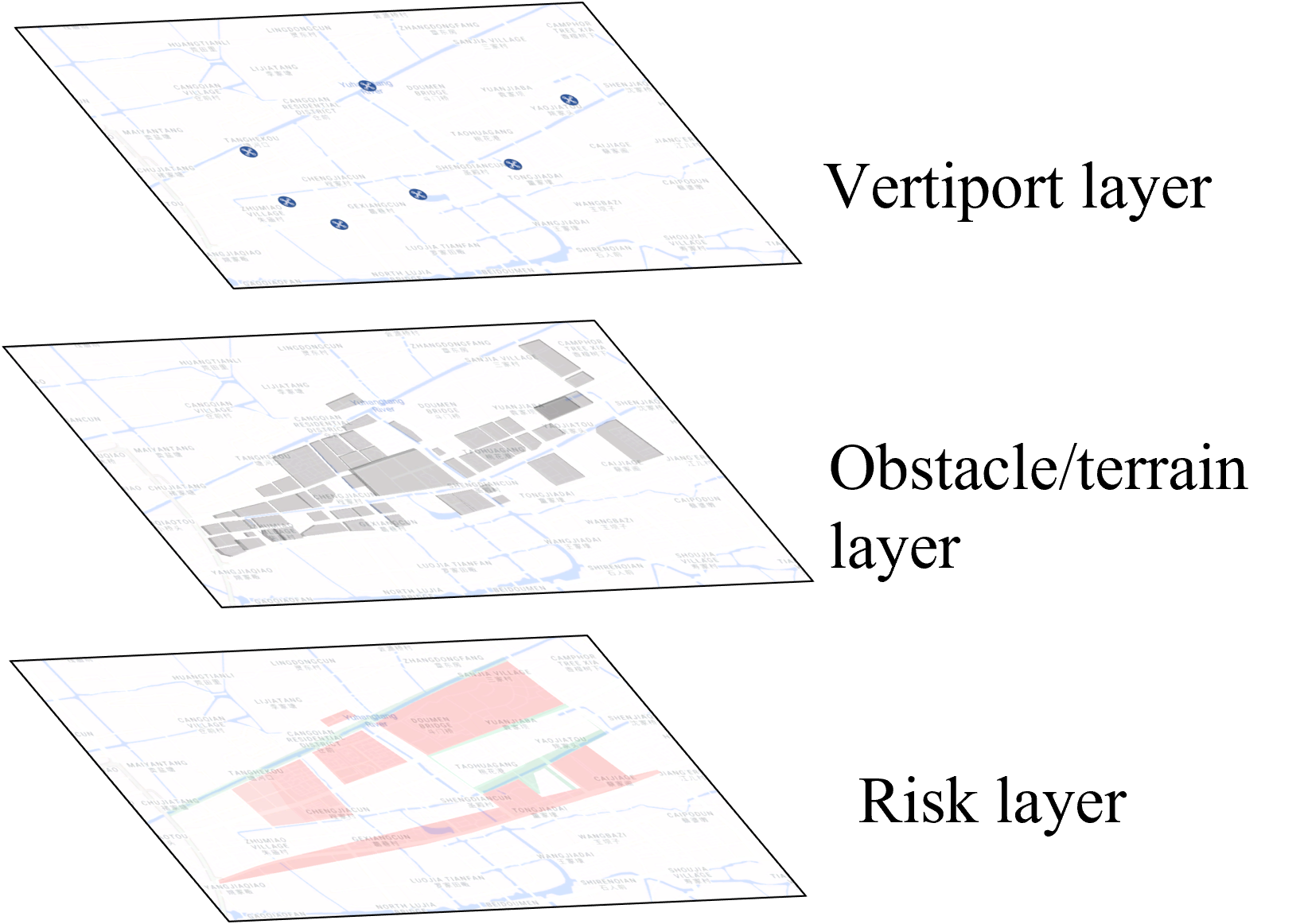}\\
  \caption{Environment layers}
\label{fig7}
\end{figure}

The environment includes ground risk, obstacles, terrains, and vertiports. They are processed as different layers, as shown in Figure \ref{fig7}. These layers are used to generate attribute values for each cell in the next discretization process. The risk layer specifies areas with high/low risks. It is calculated based on many factors, e.g., population density, sheltering factor, critical infrastructures, noise impact, public acceptance, etc. The calculation is outside the scope of this work. The obstacle/terrain layer includes obstacles like buildings, trees, no-fly zones, and terrains like flat ground and mountains. Drones should avoid any collision with obstacles and terrains. The vertiport layer provides taking off and landing points for drones. 

The space discretization process first discretizes the environment into 3D cubic grid cells, then it determines the attributes of each cell based on environment layers. Each environment layer determines the value of an attribute of a cell. The vertiport layer determines whether a cell is an origin/destination vertiport or not, the obstacle/terrain layer determines whether a cell is passable or not, and the risk layer determines what is the ground risk level associated with each cell. These grid cells compose the grid-graph $G(V, E)$ for the following network planning.

\subsection{Individual Route Planning: Extended Theta* with Space Costs}

This module aims to find a conflict-free path for each OD pair that minimizes drones' energy consumption, the potential risk to ground, and airspace occupancy. The underlying algorithm for this module is referred to as \textit{extended Theta* with space cost} in this paper. The pseudo-code of this algorithm is shown in Algorithm \ref{alg2}. The basic idea of this algorithm is explained below.

\begin{figure}[ht]
\centering
  \includegraphics[width=5cm]{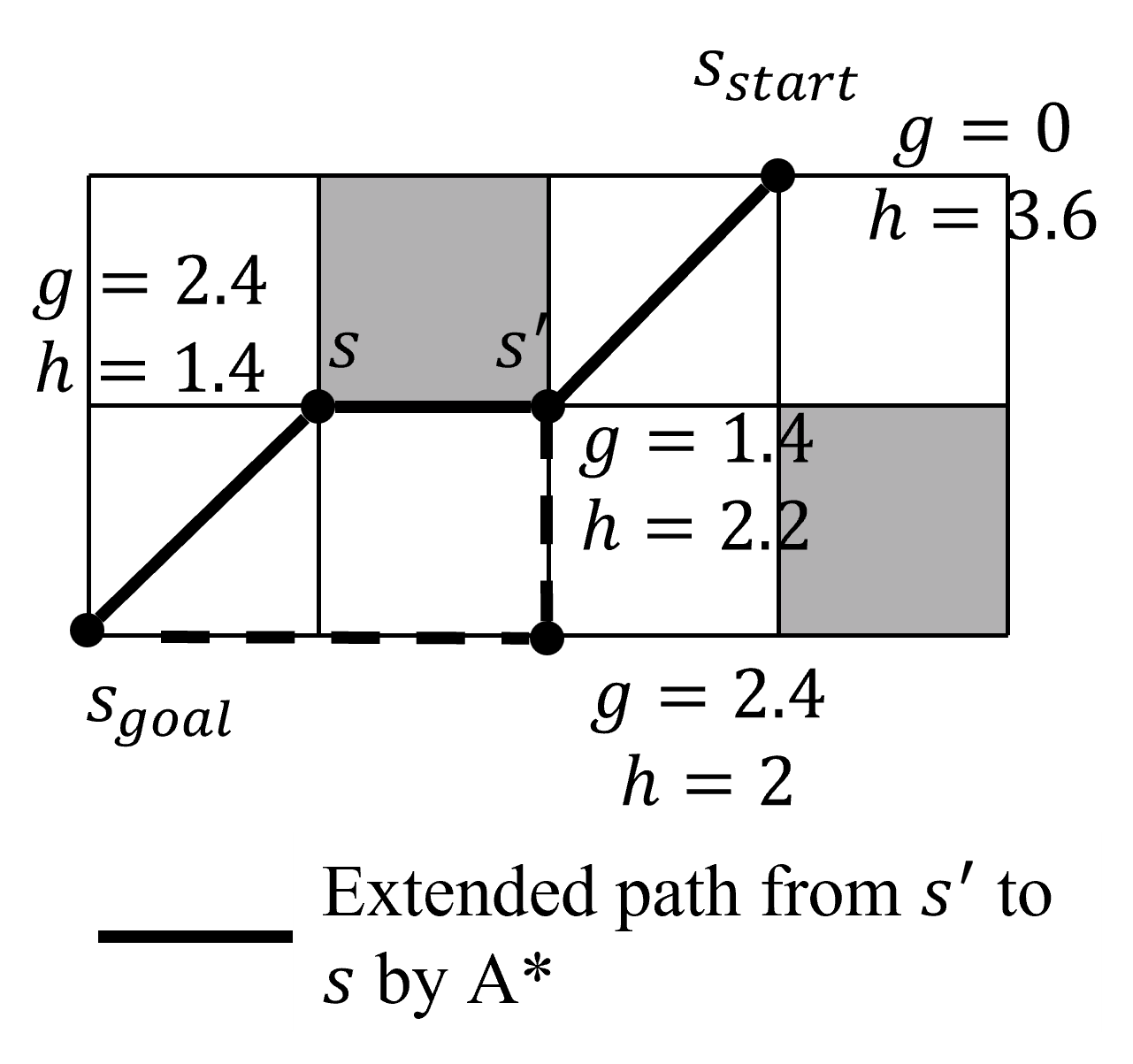}\\
  \caption{Illustration of A* path finding}
\label{fig8}
\end{figure}

The proposed path planning algorithm, extended Theta* with space cost, is developed based on the most commonly used graph-based search algorithms, A* \citep{Hart1968} and its variant Theta* \citep{Daniel2010}. In both A* and Theta*, the algorithm searches a path from a start node $s_{start}$ to a destination node $s_{goal}$ that has minimum total cost $g(s_{start},s_{goal})$. In the path searching process, the algorithm starts from the start node $s_{start}$, iteratively explores a set of nodes and determine which one would be the best as the next node to generate a path in the end. Instead of searching and evaluating all nodes in the graph, a heuristic is used to guide the algorithm to extend nodes towards the destination. The heuristic is based on the cost required to extend the path all the way to the goal, which is estimated by the direct distance from the node to the goal. If a node has a higher heuristic cost, it will be less likely to be included in the path. Specifically, in the search process of a path, to extend a node $s$ from a node $s^{'}$ towards the destination node, A* minimizes a cost function $f(s_{start},s)$, which is based on the total cost of the path $g(s_{start},s)$ and the heuristic $h(s,s_{goal})$, i.e.,
\begin{equation}
f(s_{start},s)=g(s_{start},s)+h(s,s_{goal}).
\label{eq1}
\end{equation}
An illustration for the A* path finding is shown in Figure \ref{fig8}. 

\FloatBarrier

In this paper, we modify the cost function $g$ to design each route that minimizes drones' energy consumption, the potential risk to ground, and airspace occupancy, as shown in E.q. (\ref{eq2}).
\begin{equation}
g(s_{start},s_{goal})=o(s_{start},s_{goal}) + \omega_r r(s_{start},s_{goal})+ \omega_p p(s_{start},s_{goal}).
\label{eq2}
\end{equation}
In this formula, $o(s_{start},s_{goal})$, $r(s_{start},s_{goal})$, and $p(s_{start},s_{goal})$ are operational cost, risk cost, and space cost; $\omega_r$ and $\omega_p$ are weight coefficients. 
Aside from these cost functions, a few other operational constraints might affect the route network design, such as public acceptance, noise ordinances, air traffic restrictions, etc. These operational constraints can be considered using the cost factors in a similar way. 
$\omega_r$ and $\omega_p$ in E.q. (\ref{eq2}) reflect the relative importance of risks and airspace utilization in comparison with drone flying time and energy consumption. The values can be determined by the relative monetary cost of the different categories. For example, the operational cost can be estimated by the time value of the drone delivery service and the cost of energy. The economic cost of a drone crash on ground can be estimated to gauge the risk weight value. The weight value of airspace utilization can be calculated based on the airspace usage charges or other urban airspace regulations to be established in the future.  

The operational cost captures flying distance and the energy consumption for drone operations, including traversing a distance, turning, and climbing/descending. For a set of path segments $\{\mathbi{l}_i\}_{i=1}^n$ that connects $s_{start}$ and $s_{goal}$, the operational cost is calculated as
\begin{equation}
o(s_{start},s_{goal})=traversal(s_{start},s_{goal}) + turning(s_{start},s_{goal}) + climbing(s_{start},s_{goal})+descending(s_{start},s_{goal}),
\label{eq3}
\end{equation}
where
\begin{equation}
traversal(s_{start},s_{goal})=\sum_{i=1}^{n} \left\Vert \mathbi{l}_i\right\Vert,
\label{eq4}
\end{equation}
\begin{equation}
turning(s_{start},s_{goal})=\lambda_{turning} \sum_{i=1}^{n-1} \left\vert arccos(\frac{\mathbi{l}_{i}\cdot \mathbi{l}_{i+1}}{\left\Vert \mathbi{l}_{i}\right\Vert \left\Vert \mathbi{l}_{i+1}\right\Vert})\right\vert,
\label{eq5}
\end{equation}
let $\mathbi{n}=[0, 0, 1]^{T}$, then
\begin{equation}
climbing(s_{start},s_{goal})=\lambda_{climbing} \sum_{i=1}^{n} max(arcsin(\frac{\mathbi{l}_{i}\cdot \mathbi{n}}{\left\Vert \mathbi{l}_{i} \right\Vert }),0)\left\Vert \mathbi{l}_{i} \right\Vert,
\label{eq6}
\end{equation}
\begin{equation}
descending(s_{start},s_{goal})=\lambda_{descending} \sum_{i=1}^{n} max(-arcsin(\frac{\mathbi{l}_{i}\cdot \mathbi{n}}{\left\Vert \mathbi{l}_{i} \right\Vert }),0)\left\Vert \mathbi{l}_{i} \right\Vert.
\label{eq7}
\end{equation}
Here $\lambda_{turning}, \lambda_{climbing}, \lambda_{descending}$ are coefficients to normalize energy consumption for different drone operations.

The risk cost captures the potential risk to the ground, it reflects various risk levels by accumulating the risks involved in passing through the cells. It is calculated as
\begin{equation}
r(s_{start},s_{goal})=\lambda_r \sum_{s_{start}}^{s_{goal}} \theta_{risk},
\label{eq8}
\end{equation}
where $\theta_{risk}$ indicates the risk level of an area. Areas that are densely populated or with critical infrastructures have higher risk levels, with $\theta_{risk}>1$; areas that are not populated and with less ground impact concerns have lower risk levels, with $0<\theta_{risk}<1$; most areas are set to have a normal risk level, with $\theta_{risk}=1$. $\lambda_r$ is a scaling factor for raw risk cost, the calculation of it is in Appendix A. 

\begin{figure}[ht]
\centering
\subfigure[buffer zones do not overlap]{
   \includegraphics[scale =0.6] {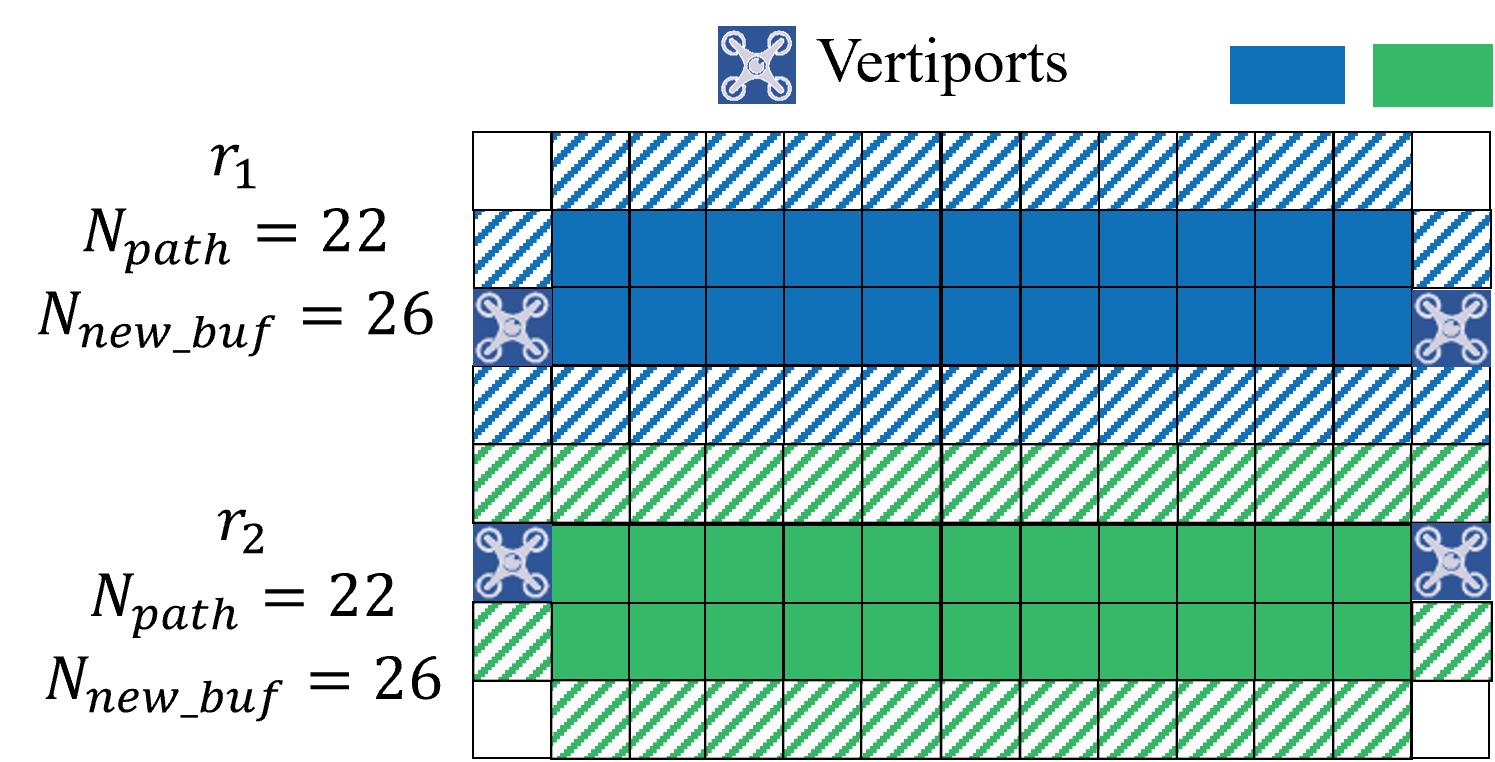}
   \label{subfig91}
 }
 \subfigure[buffer zones overlap]{
   \includegraphics[scale =0.6] {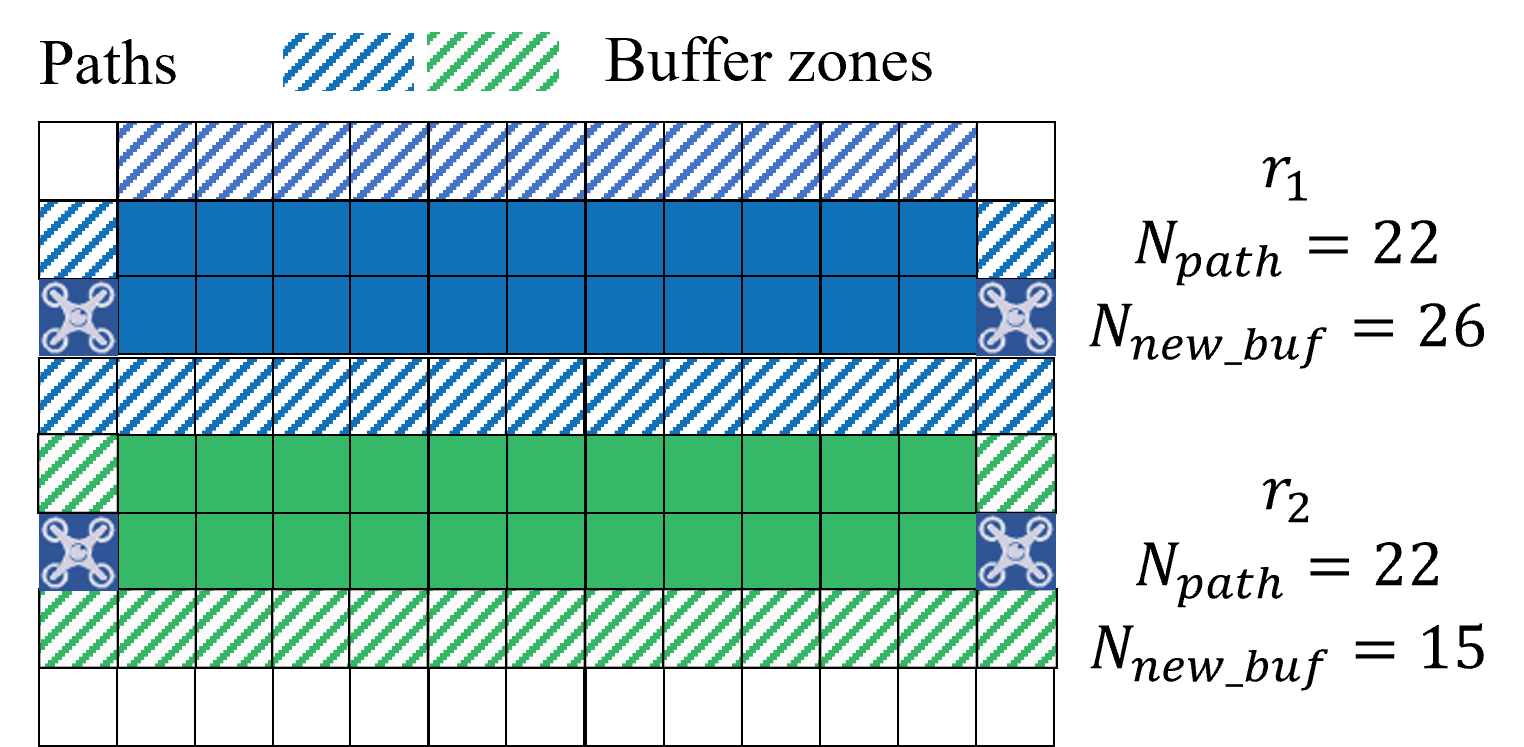}
   \label{subfig92}
 }
  \caption{An illustration for overlapping buffer zones (top view)}
\label{fig9}
\end{figure}

The space cost function encourages bundled paths and overlapped buffer zones to improve airspace utilization. An illustration is shown in Figure \ref{fig9}. The size of buffer zones in Figure \ref{fig9}(a) and Figure \ref{fig9}(b) are the same and equal to the required minimum separation. In Figure \ref{fig9}(a), the buffer zones of two aligned paths are not overlapped and the separation between the paths is twice the required minimum separation; in Figure \ref{fig9}(b), the buffer zones are overlapped and the separation between the paths is exactly the required minimum separation. For a path between cell $s_{start}$ and cell $s_{goal}$, the space cost item $p(s_{start},s_{goal})$ measures the marginal volume of occupied airspace, i.e., 
\begin{equation}
p(s_{start},s_{goal})=\lambda_p N(s_{start},s_{goal}), 
\label{eq9}
\end{equation}
here $\lambda_p$ is a scaling factor for raw space cost, the calculation of it is in Appendix A, $N(s_{start},s_{goal})$ is calculated by
\begin{equation}
N(s_{start},s_{goal})=N_{path} (s_{start},s_{goal})+N_{new\_{buf}} (s_{start},s_{goal}),
\label{eq10}
\end{equation}
where $N_{path}$ is the increased number of path cells by adding a path between cell $s_{start}$ and cell $s_{goal}$ to the existing network, while $N_{new\_{buf}}$ is the increased number of buffer zone cells by adding a path between cell $s_{start}$ and cell $s_{goal}$ to the existing network. Taking Figure \ref{fig9} as an example, buffer zones in Figure \ref{subfig91} do not overlap, so all buffer zone cells are newly introduced. For the situation in Figure \ref{subfig92}, when planning a path $r_2$ after path $r_1$ has been planned, some of $r_2$'s buffer zone cells overlapped with the buffer zones of $r_1$. These buffer zone cells will not be counted again in $N_{new\_{buf}}$ for $r_2$. After the addition of the space cost function, the total cost reduces but the length of paths increases, but this is not an issue. As a drone delivery operator, the cost is associated with the direct drone operations as well as the airspace usage charges, similar to the airline operations.

To ensure the effectiveness of the space cost function for the purpose of reducing total costs, the relative value of the space cost weight coefficient $\omega_p$ in relation to other cost coefficients needs to be carefully calibrated to reflect the actual operational cost impact of airspace usage charges. For example, when the space cost weight coefficient $\omega_p$ is small, the space cost has little impact on route density. As $\omega_p$ increases, the algorithm starts to reduce large space costs at the expense of a small increase in other costs. When $\omega_p$ is large, the algorithm generates a network where paths are heavily bundled to have more overlapped buffer zones, but the paths may pass through areas with high operational cost and risk cost. 

\begin{sloppypar}
\begin{algorithm}[H]
\algorithmfootnote{The proposed space cost function is highlighted as red in the Algorithm.}
\begin{multicols}{2}
    \renewcommand{\thealgocf}{2}
    \SetAlgoLined
    \DontPrintSemicolon
    \SetKwFunction{ComputeCost}{ComputeCost}
    \SetKwFunction{SpaceCost}{SpaceCost}
    \SetKwFunction{Main}{Main}
    \SetKwFunction{UpdateVertex}{UpdateVertex}
    \SetKwProg{Fn}{}{:}{}
    \Fn{\Main{}}{
    $open:=closed:=\emptyset$\;
    $g(s_{start},s_{start}):=0$\;
    $parent(s_{start}):=s_{start}$\;
    $open.Insert(s_{start},g(s_{start},$\;
    $s_{start})+h(s_{start},s_{goal}))$\;
    \While{$open \neq \emptyset$}{
    $s^{'}:=open.Pop()$\;
    [SetVertex($s^{'}$)]\;
    \If{$s^{'}=s_{goal}$}{
    \Return "path found"
    }
    $closed := closed \cup \{s^{'}\}$\;
    \ForEach{$s \in nghbr_{vis}(s^{'})$}{
    \If{$s \notin closed $}{
    \If{$s \notin open$}{
    $g(s_{start},s):=\infty$\;
    $parent(s):=NULL$\;
    }
    UpdateVertex($s^{'},s$)
    }
    }
    }
    \Return "no path found"
    }

    \Fn{\UpdateVertex{$s^{'},s$}}{
	$g_{old}:=g(s_{start},s)$\;
    ComputeCost($s^{'},s$)\;
    \If{$g(s_{start},s)<g_{old}$}{
    \If{$s \in open$}{
    $open.Remove(s)$
    }
    $open.Insert(s,g(s_{start},s)+h(s,s_{goal}))$
    }
    }
    \color{red}
    \Fn{\SpaceCost{$s^{'},s$}}{
    	/* calculate space cost \;
    	traveling from $s^{'}$ to $s$ */\;
    	$N(s^{'},s)=N_{path}(s^{'},s)+N_{new\_pro}(s^{'},s)$\;
    	$p(s^{'},s)=\lambda_pN(s^{'},s)$\;
    	\Return $p(s^{'},s)$
    }
    \color{black}
    \Fn{\ComputeCost{$s^{'},s$}}{
    \uIf{$lineofsight(parent(s^{'}),s)$}{
    /* Path 2 */\;
    $\color{red}p(parent(s^{'}),s)=SpaceCost(parent(s^{'}),s)$\;
    $g(parent(s^{'}),s)=o(parent(s^{'}),s)+\omega_r r(parent(s^{'}),s)+\color{red} \omega_p p(parent(s^{'}),s)$\;
    \If{$g(s_{start},parent(s^{'}))+g(parent(s^{'}),s)<g(s_{start},s)$}{
    $parent(s):=parent(s^{'})$\;
    $g(s_{start},s) : = g(s_{start},parent(s^{'}))$ \;$+g(parent(s^{'}),s)$\;
    }
    }
    \Else{
        /* Path 1 */\;
        $\color{red}p(s^{'},s)=SpaceCost(s^{'},s)$\;
        $g(s^{'},s)=o(s^{'},s)+\omega_r r(s^{'},s)+\color{red} \omega_p p(s^{'},s)$\;
    \If{$g(s_{start},s^{'})+g(s^{'},s)<g(s_{start},s)$}{
    $parent(s):=s^{'}$\;
    $g(s_{start},s) : = g(s_{start},s^{'})+g(s^{'},s)$\;
    }
    }
    }
    \caption{Extended Theta* with Space Cost}
    \label{alg2}
\end{multicols}
\end{algorithm}
\end{sloppypar}

\subsection{Route Prioritization and Network Planning}

This module specifies the prioritization framework based on a set of prioritization rules to rank the importance of routes and decouple the network planning problem into sequential planning. As discussed in the introduction, the network-planning problem is an NP-hard problem. An optimal solution of a single path occupies certain airspace, which may force other paths to detour. Thus, a change of one path affects all other paths. In this paper, we use a simple but effective strategy to decouple the network planning problem into a set of single-path planning problems - plan the paths one by one from the most important to the least important. 

\begin{figure}[ht]
\centering
  \includegraphics[width=15cm]{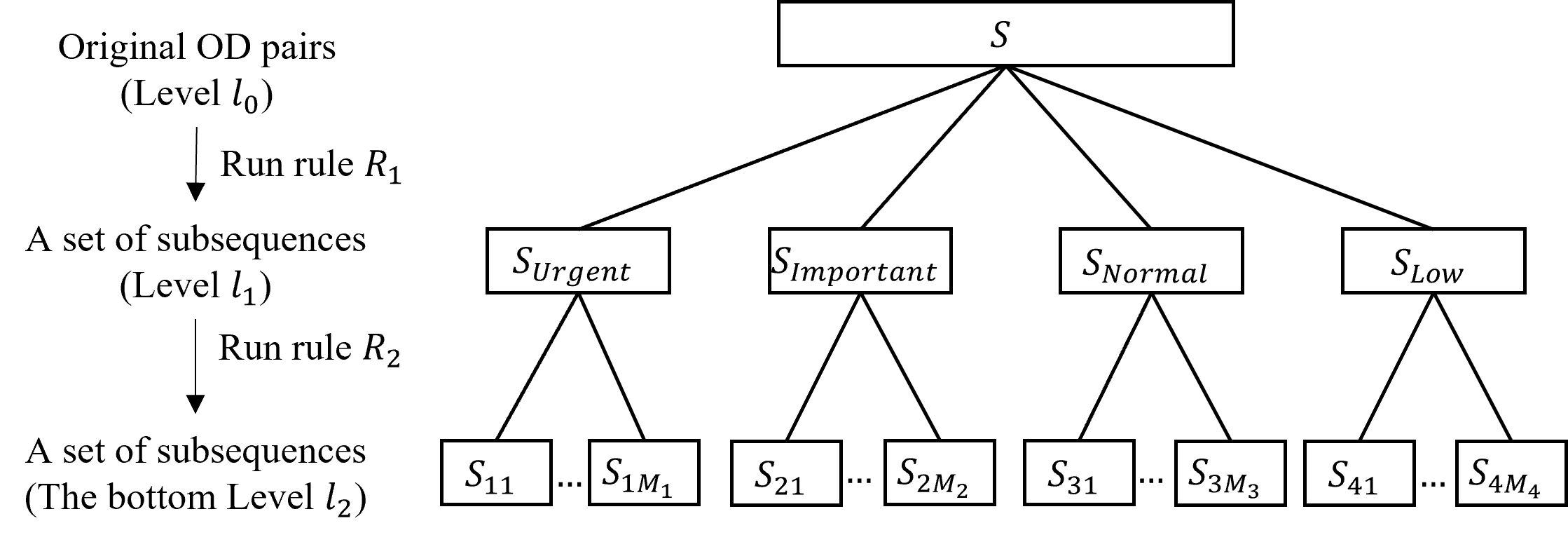}\\
  \caption{Illustration of the prioritization framework}
\label{fig10}
\end{figure}

In this module, a prioritization structure with example rules is proposed. These rules can be changed depending on specific business considerations. The prioritization structure is shown in Figure \ref{fig10}. There can be multiple levels to prioritize the paths. Level $l_0$ has the sequence of original OD pairs and it is grouped into a series of subsequences by the most important priority rule, $R_1$. Then the subsequences on Level $l_1$ can be further grouped into subsequences by the second important priority rule, $R_2$. The paths in the subsequences on the bottom level are randomly shuffled to find the optimal sequence that generates the best network performance.

\FloatBarrier

To give a more detailed example, suppose we have two priority rules as specified below:
\begin{equation}
R_1=\{PL\},
\label{eq11}
\end{equation}
\begin{equation}
R_2=\{v,\varepsilon_v \},
\label{eq12}
\end{equation}
where $PL$ denotes the urgency level of paths, detailed as $\{Urgent,Important,Normal,Low\}$ ; $v$ is the expected profit value of each path and $\varepsilon_v$ is segment threshold of $v$. More priority rules can be taken to get more subsequences.

\begin{figure}[ht]
\centering
  \includegraphics[width=12cm]{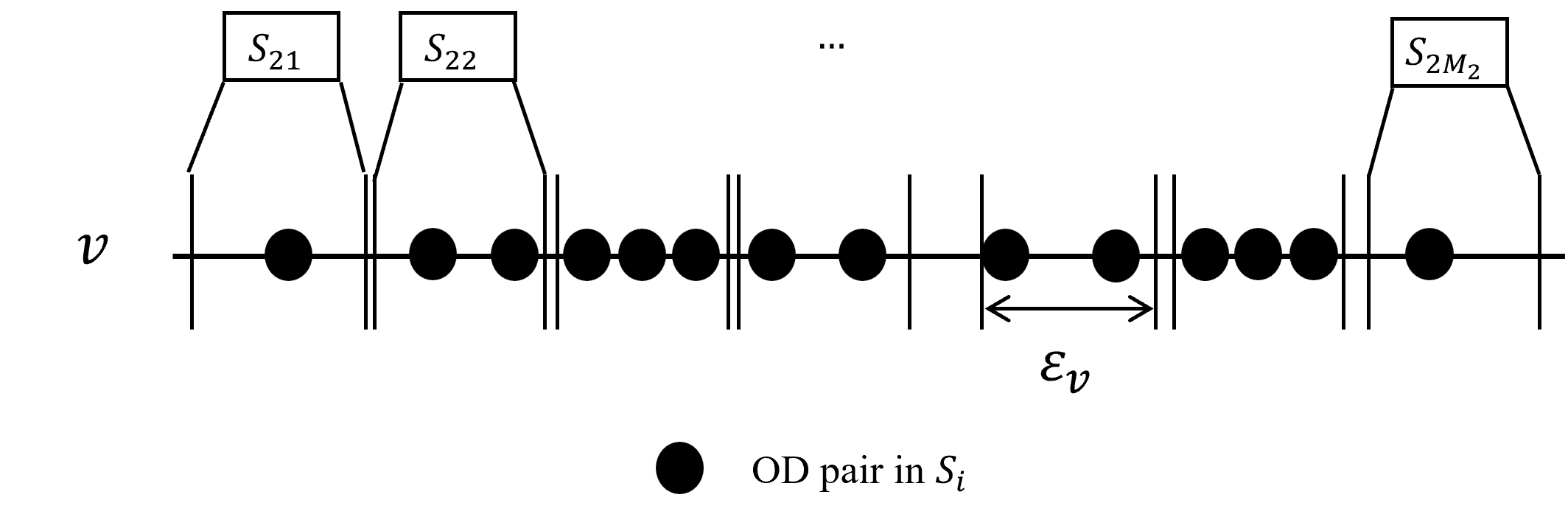}\\
  \caption{Illustration of the segmentation and shuffle in $R_2$}
\label{fig11}
\end{figure}

$R_1$ generates four subsequences $\{S_i\}_{i=1}^{4}$ directly based on urgency levels in $PL$. These subsequences are $(S_{Urgent},$ $S_{Important},S_{Normal},S_{Low})$, each of them represents an urgency level. $R_2$ sorts paths within each subsequence $S_i$ in $l_1$ by the expected profit value $v$ (in descending order), and generates a new set of subsequences $\{S_{ij}\}_{j=1}^{M_i}$, where $M_i$ is the number of subsequences in $S_i$. $\varepsilon_v$ divides each $S_i$ into a set of subsequences $\{S_{ij}\}_{j=1}^{M_i}$ based on the expected profit value distribution (Figure \ref{fig11}), similar to generating clusters based on density in DBSCAN \citep{Ester1996}. Given a set of paths with different $v$, OD pairs are grouped together with similar $v$ values. Each $S_{ij}$ satisfies the following property:
\begin{equation}
|v_r-v_{r^{'}}|\leq\varepsilon_v, \forall r,r^{'} \in S_{ij}.
\label{eq13}
\end{equation}
Paths in the subsequences on the bottom level are randomly shuffled to find the optimal sequence that generates the best network performance. 

\FloatBarrier

To balance between optimality and computational time, we randomly generate $K$ sequences following the above priority structure, where $K$ is a parameter that one can adjust. It can be adjusted in the range $[1,S_g]$,
\begin{equation}
K\in \mathbb{Z}: 1\leq K \leq S_g.
\label{eq14}
\end{equation}

\begin{equation}
S_g=\prod_{i=1, 2, 3, 4}\prod_{j = 1, 2, ..., M_i} n(S_{ij})!,
\label{eq15}
\end{equation}
where $n(S_{ij})$ is the number of OD pairs in sub-sequence $S_{ij}$ , $n(S_{ij})!$ is the number of permutations in sub-sequence $S_{ij}$,  and $S_g$ is the number of all possible ordered arrangements of OD pairs satisfying $R_1$ and $R_2$. 

The complete strategy for determining route sequence order is shown in Algorithm \ref{alg3}. If there are two networks that generate the same performance in terms of the total costs, the algorithm will generate both networks and their associated costs in different categories. It's up to the user to select which one to use based on the information of different cost categories and other considerations not captured in the model.

\begin{algorithm}[H]
    \renewcommand{\thealgocf}{3}
    \SetAlgoLined
    \DontPrintSemicolon
    \SetKwInOut{Input}{Input}
    \SetKwInOut{Output}{Output}
    \SetKwInOut{Initialize}{Initialize}
    \Input{Original OD pairs $\{OD_i\}_{i=1}^N$}
    \Output{Ordered OD pairs $\{OD_{i}^{'}\}_{i=1}^N$}
    /* Rule 1: get $\{S_{i}\}^{4}_{i=1}$, i.e., $(S_{Urgent},$ $S_{Important},S_{Normal},S_{Low})$ */\;
    $\{S_{i}\}^{4}_{i=1}\gets \emptyset$\;
    \For{$r$ in $\{OD_i\}_{i=1}^N$}{
    	$S_l.append(r)$ if $r.PL==l$\;
     }
    $\{S_{i}\}^{4}_{i=1} \gets \{S_l\}$\;\;
    /* Rule 2: get $\{\{S_{ij}\}_{j=1}^{M_i}\}^{4}_{i=1}$*/\;
   	\For{$S_i$ in $\{S_{i}\}^{4}_{i=1}$}{
    	Sort $S_i$ in descending order of $v$;\;
      	$\{S_{ij}\}_{j=1}^{M_i} \gets$ subgroups from DBSCAN with $\varepsilon_v$ in $S_i$;\;
      	\For{$S_{ij}$ in $\{S_{ij}\}_{k=1}^{M_{i}}$}{
      		$S_{ij} \gets $ random shuffled $S_{ij}$;\;
      	}
    }
\;
     $\{OD_{i}^{'}\}_{i=1}^N \longleftarrow \{\{S_{ij}\}_{j=1}^{M_{i}}\}^{4}_{i=1}$;\;
     return $\{OD_{i}^{'}\}_{i=1}^N$
    \caption{Route Prioritization}
    \label{alg3}
\end{algorithm}

\section{Testing}
In this section, we first use an illustrative scenario to show how the algorithm works. Then we compare the proposed algorithm with other path finding algorithms in a test scenario. After that, the proposed algorithm is applied to a real-world scenario in Hangzhou, China. In this real-world scenario, we show a set of sensitivity analyses on algorithm parameters and provide an empirical analysis of computational time. 
\subsection{Illustration with a Toy Example}
\begin{figure}[ht]
\centering
\subfigure[Our method without space cost]{
   \includegraphics[scale =0.5] {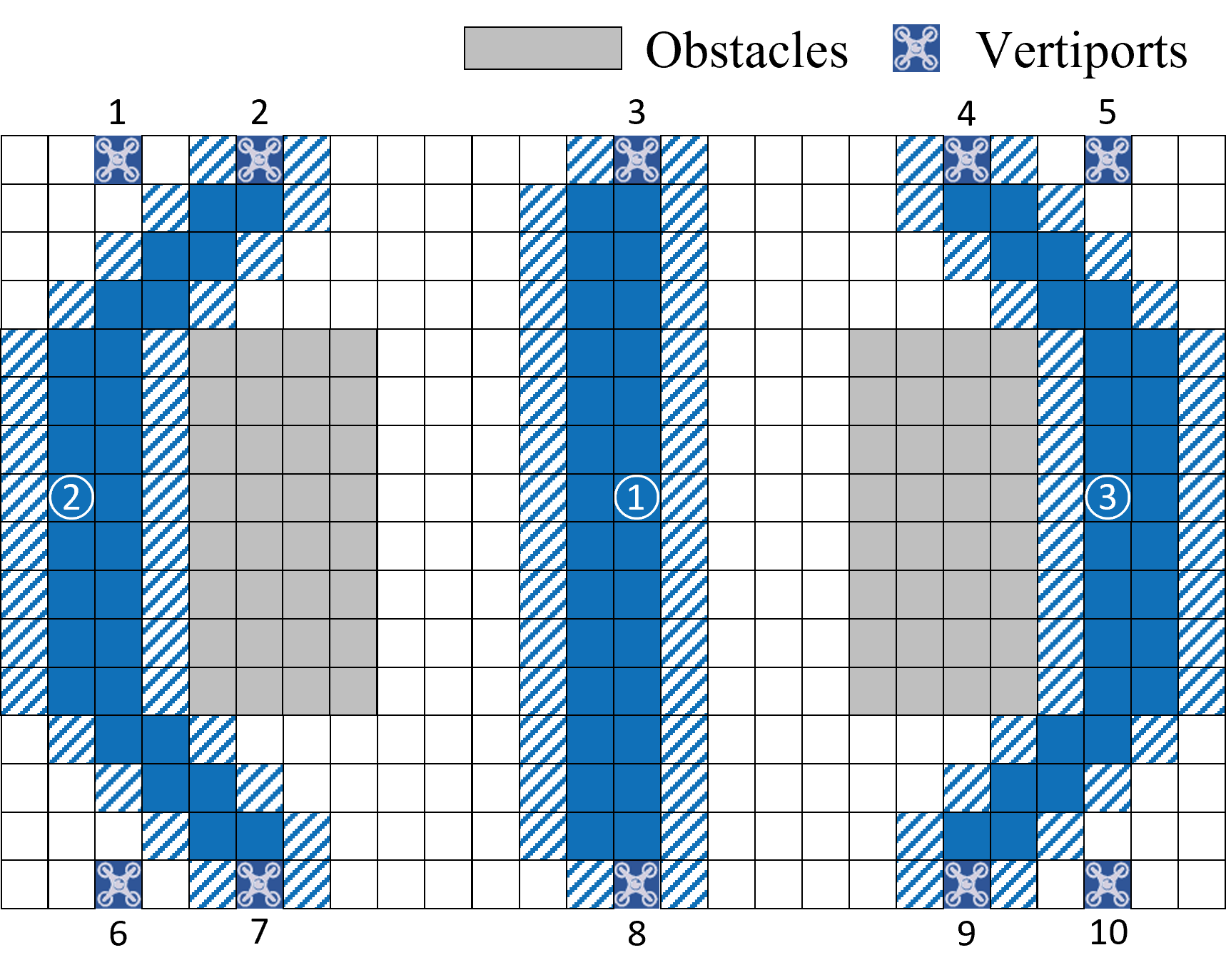}
   \label{subfig121}
 }
 \subfigure[Our method with space cost]{
   \includegraphics[scale =0.5] {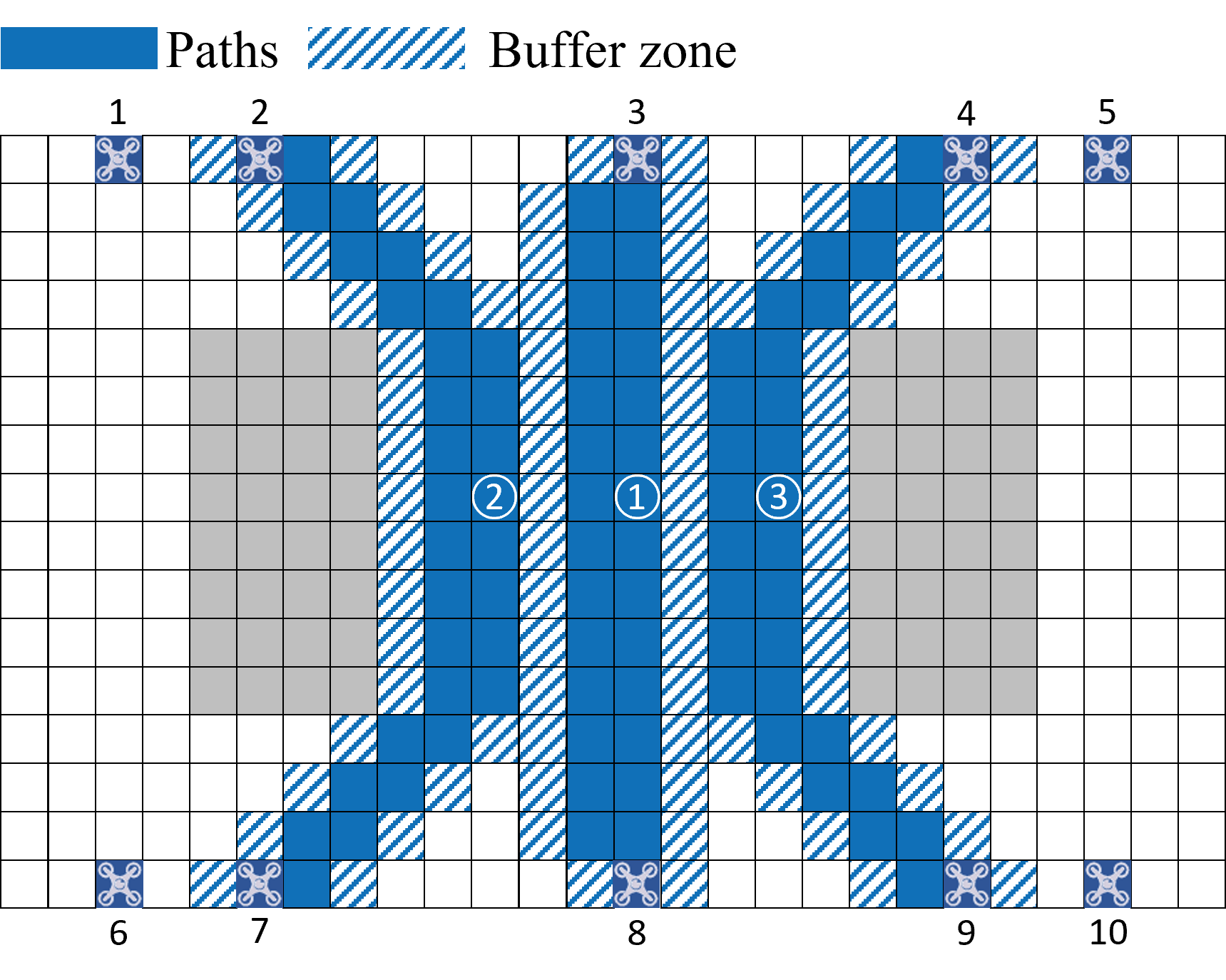}
   \label{subfig122}
 }
 \caption{ A test case (top view) for network planning (a) without space cost and (b) with space cost}
\label{fig12}
\end{figure}

In this section, we use a toy example to show how the space cost function improves airspace utilization for the network design. We apply the \textit{our method with space cost} algorithm and the \textit{our method without space cost} algorithm on the toy example shown in Figure \ref{fig12}. In this example, there are two symmetric obstacles and 10 vertiports (id 1-10).

In Experiment 1, both algorithms are applied to find three paths from upper vertiports to lower vertiports (id 3-8, 2-7, 4-9). The results are shown in Figure \ref{fig12}. If the space cost function is not added to the algorithm, three paths (\textcircled{1}, \textcircled{2}, \textcircled{3}) will appear on the different sides of the obstacles to achieve a shorter distance, as shown in Figure \ref{subfig121}; if the space cost function is added to the algorithm, three paths all appear in the middle of two obstacles and share the buffer zones, as shown in Figure \ref{subfig122}. After adding the space cost function into the algorithm, the generated network occupies fewer buffer zones (16.7\%) at the cost of slightly more path cells (4.8\%), and the total occupied airspace reduces (6.7\%), as shown in Table \ref{table1}. 
\FloatBarrier

\begin{table}[ht]
    \centering
    \caption{Summary of planning three paths in the scenario}
    \label{table1}
    \begin{tabular}{c c c c}
        \hline
          & \# of buffer zone cells & \# of path cells & Total occupied airspace \\
        \hline
        Our method without space cost & 96 & 84 & 180 \\
        Our method with space cost & 80 & 88 & 168 \\
        \hline
        Improvement after adding space cost     & 16.7\% & -4.8\% & 6.7\% \\
        \hline
    \end{tabular}
\end{table}
\FloatBarrier

\begin{figure}[ht]
\centering
  \includegraphics[scale =0.5]{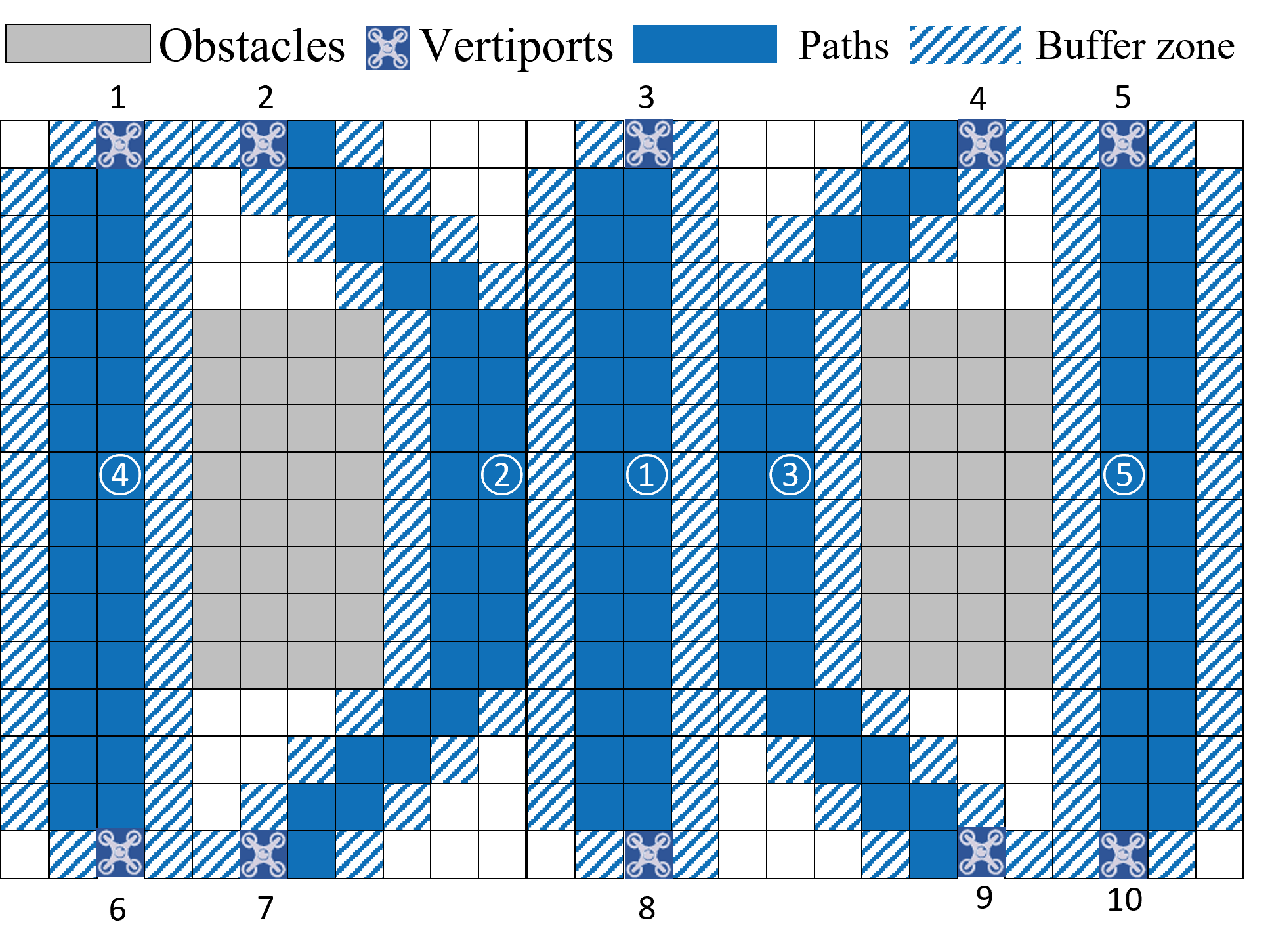}\\
  \caption{A test case (top view) of network planning with 5 paths}
\label{fig13}
\end{figure}

In Experiment 2, both algorithms are applied to find five paths from upper vertiports to lower vertiports (id 3-8, 2-7, 4-9, 1-6, 5-10). The results are shown in Figure \ref{fig13}. When not adding the space cost function, the method fails to find the paths \textcircled{4} and \textcircled{5} because their traversable airspaces are blocked by the buffer zones of the paths \textcircled{2} and \textcircled{3}. By contrast, when adding the space cost function, the method successfully generates the extra paths (\textcircled{4}, \textcircled{5}).

\FloatBarrier

\subsection{Comparisons with Other Algorithms on a 2D Scenario}
\begin{figure}[ht]
\centering
  \includegraphics[width=5cm]{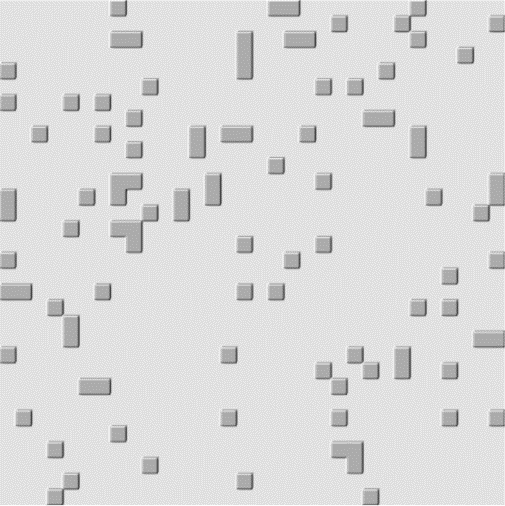}\\
  \caption{The 2D standard test scenario \citep{sturtevant2012benchmarks}}
\label{fig14}
\end{figure}

In this section, we compare the proposed sequential route network planning method with two kinds of state-of-the-art MPF algorithms on a 2D standard test scenario: one is a two-level-based optimal method - conflict-based search (CBS) \citep{Sharon2015}, and one is a rule-based sub-optimal method - Push-and-Spin \citep{Alotaibi2018}. The scenario is shown in Figure \ref{fig14}. This scenario \citep{sturtevant2012benchmarks} contains 65536 pixels/cells, and 6528 cells are inaccessible. Multiple OD pairs need to be planned in this experiment. The origins are located in the lower-left corner and the destinations are located in the upper-right corner.

\FloatBarrier

Results show that the proposed method is capable to solve large scenarios within a reasonable amount of time. The comparisons are shown in Table \ref{table2}. Both Push-and-Spin and the proposed method find paths quickly, while CBS cannot find paths in a reasonable time as the number of routes increases. The proposed method also shows an advantage in airspace utilization compared to CBS and Push-and-Spin. The paths generated by these algorithms are shown in Figure \ref{fig15}. 

\begin{table}[h]
    \centering
    \caption{Route network planning results of three methods in the 2D standard test scenario}
    \label{table2}
    \begin{threeparttable}
    \resizebox{\textwidth}{18mm}{
    \begin{tabular}{p{3cm} c c c c c c c c c c c c c c c c c c c c}
        \hline
        \# of routes & 1 & 2 & 3 & 4 & 5 & 6 & 7 & 8 & 9 & 10 & 11 & 12 & 13 & 14 & 15 & 16 & 17 & 18 & 19 & 20\\
        \hline
        \multicolumn{7}{l}{\textbf{CBS}} \\
        $sum \ of \ distance $ & 245 & 477 & 703 & 956 & 1203 & 1443 & 1694 & 1965 & 2263 & - & - & - & - & - & - & - & - & - & - & - \\
        \# of occupied cells & 1007 & 1923 & 2910 & 3965 & 4894 & 5918 & 6902 & 7993 & 9189 & - & - & - & - & - & - & - & - & - & - & - \\
        $time(s)$ & 0.02 & 0.03 & 1.98 & 134.04 & 1201.34 & 2029.10 & 2378.24 & 2836.76 & 3165.76 & - & - & - & - & - & - & - & - & - & - & - \\
        \hline
        \multicolumn{7}{l}{\textbf{Push-and-Spin}} \\
        $sum \ of \ distance$ & 245 & 477 & 707 & 966 & 1203 & 1466 & \# & \# & \# & \# & \# & \# & \# & \# & \# & \# & \# & \# & \# & \# \\
        \# of occupied cells & 1007 & 1923 & 2852 & 3904 & 4894 & 5877 & \# & \# & \# & \# & \# & \# & \# & \# & \# & \# & \# & \# & \# & \# \\
        $time(s)$ & 0.02 & 0.03 & 0.04 & 0.04 & 0.05 & 0.06 & \# & \# & \# & \# & \# & \# & \# & \# & \# & \# & \# & \# & \# & \# \\
        \hline
        \multicolumn{7}{l}{\textbf{The proposed method}} \\
        $sum \ of \ distance$ & 247 & 486 & 715 & 969 & 1226 & 1495 & 1734 & 1999 & 2266 & 2564 & 2893 & 3175 & 3490 & 3807 & 4087 & 4367 & 4666 & 4968 & 5316 & 5664 \\
        \# of occupied cells & 881 & 1491 & 2019 & 2771 & 3458 & 4121 & 4915 & 5699 & 6623 & 7400 & 8232 & 8949 & 9713 & 10522 & 11170 & 11867 & 12623 & 13349 & 14229 & 15311 \\
        $time(s)$ & 0.16 & 0.21 & 0.26 & 0.32 & 0.42 & 0.52 & 0.58 & 0.67 & 0.77 & 0.85 & 0.92 & 0.97 & 1.03 & 1.09 & 1.12 & 1.16 & 1.21 & 1.25 & 1.29 & 1.33 \\
        \hline
    \end{tabular}
    }
    \begin{enumerate}
    \setlength{\itemsep}{0pt}
    \setlength{\parsep}{0pt}
    \setlength{\parskip}{0pt}
    	\footnotesize
    	\item[-:] exceed the timeout limit (3600s)
    	\item[\#:] unable to find feasible solutions
    \end{enumerate}
   \end{threeparttable}
\end{table}
\FloatBarrier

\begin{figure}[h]
\centering
\subfigure[3 routes - CBS]{
   \includegraphics[width=5cm] {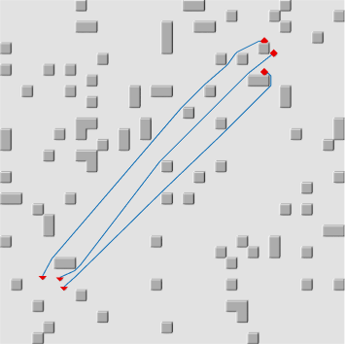}
   \label{subfig151}
 }
 \subfigure[3 routes - Push-and-Spin]{
   \includegraphics[width=5cm] {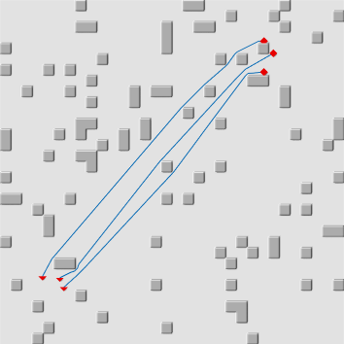}
   \label{subfig152}
 }
 \subfigure[3 routes - The proposed method]{
   \includegraphics[width=5cm] {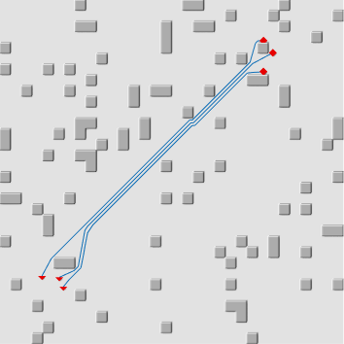}
   \label{subfig153}
 }
 \subfigure[7 routes - CBS]{
   \includegraphics[width=5cm] {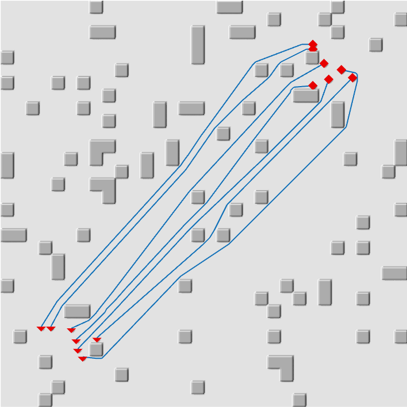}
   \label{subfig154}
 }
 \subfigure[7 routes - Push-and-Spin]{
   \includegraphics[width=5cm] {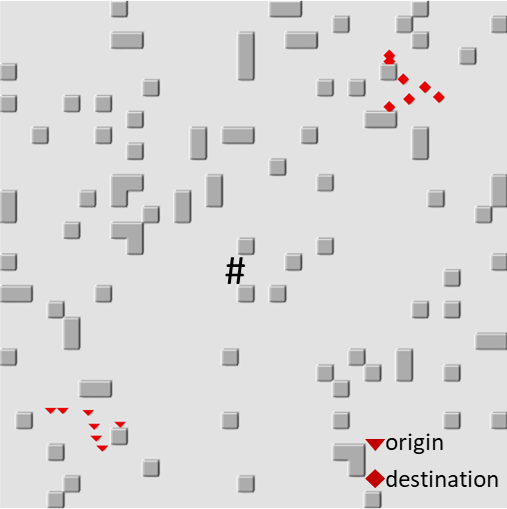}
   \label{subfig155}
 }
 \subfigure[7 routes - The proposed method]{
   \includegraphics[width=5cm] {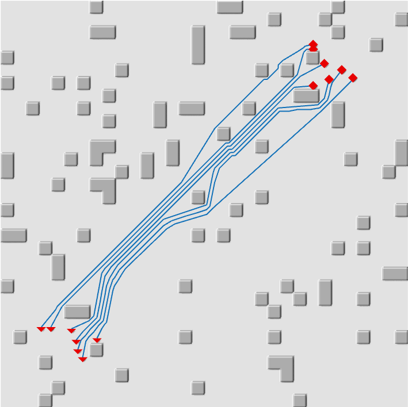}
   \label{subfig156}
 }
  \subfigure[10 routes - CBS]{
   \includegraphics[width=5cm] {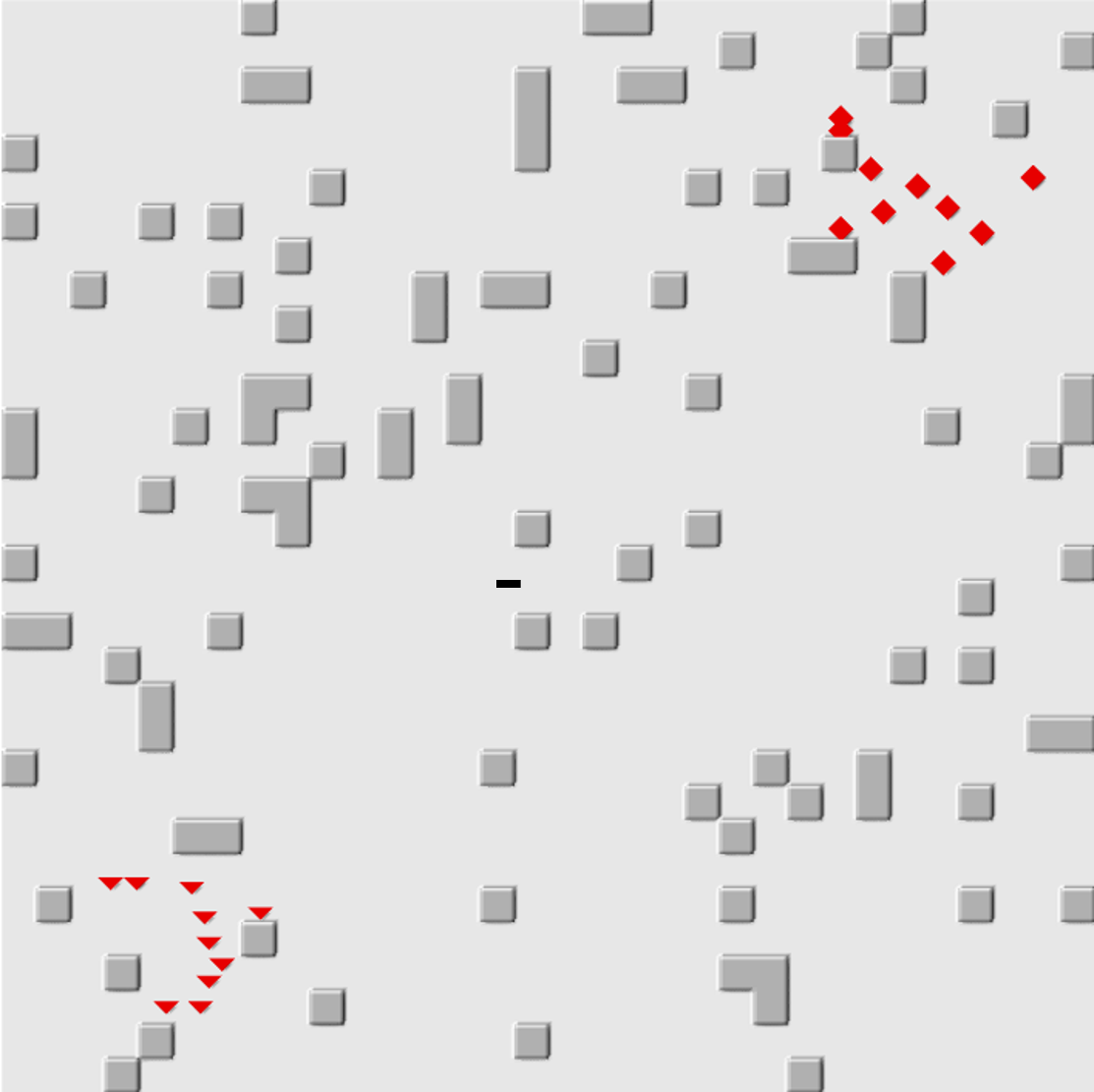}
   \label{subfig157}
 }
 \subfigure[10 routes - Push-and-Spin]{
   \includegraphics[width=5cm] {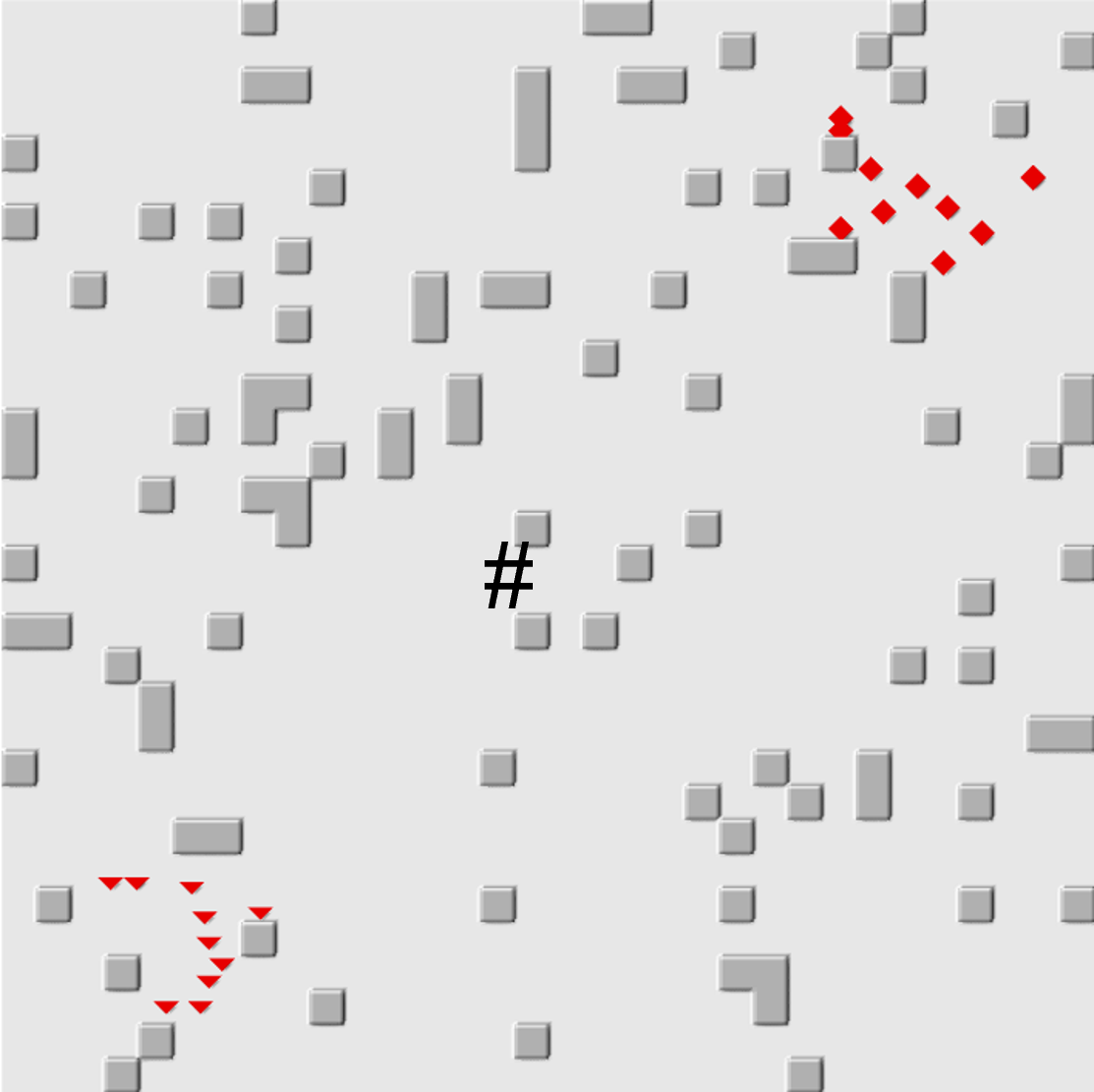}
   \label{subfig158}
 }
 \subfigure[10 routes - The proposed method]{
   \includegraphics[width=5cm] {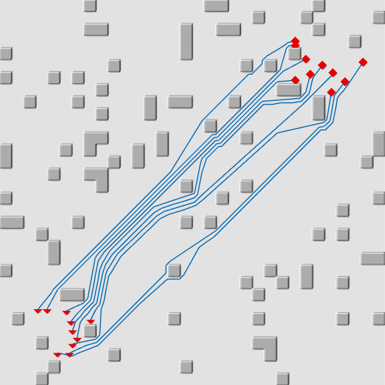}
   \label{subfig159}
 }
 \end{figure}
\FloatBarrier
\begin{figure}[h]
\centering
  \subfigure[15 routes - CBS]{
   \includegraphics[width=5cm] {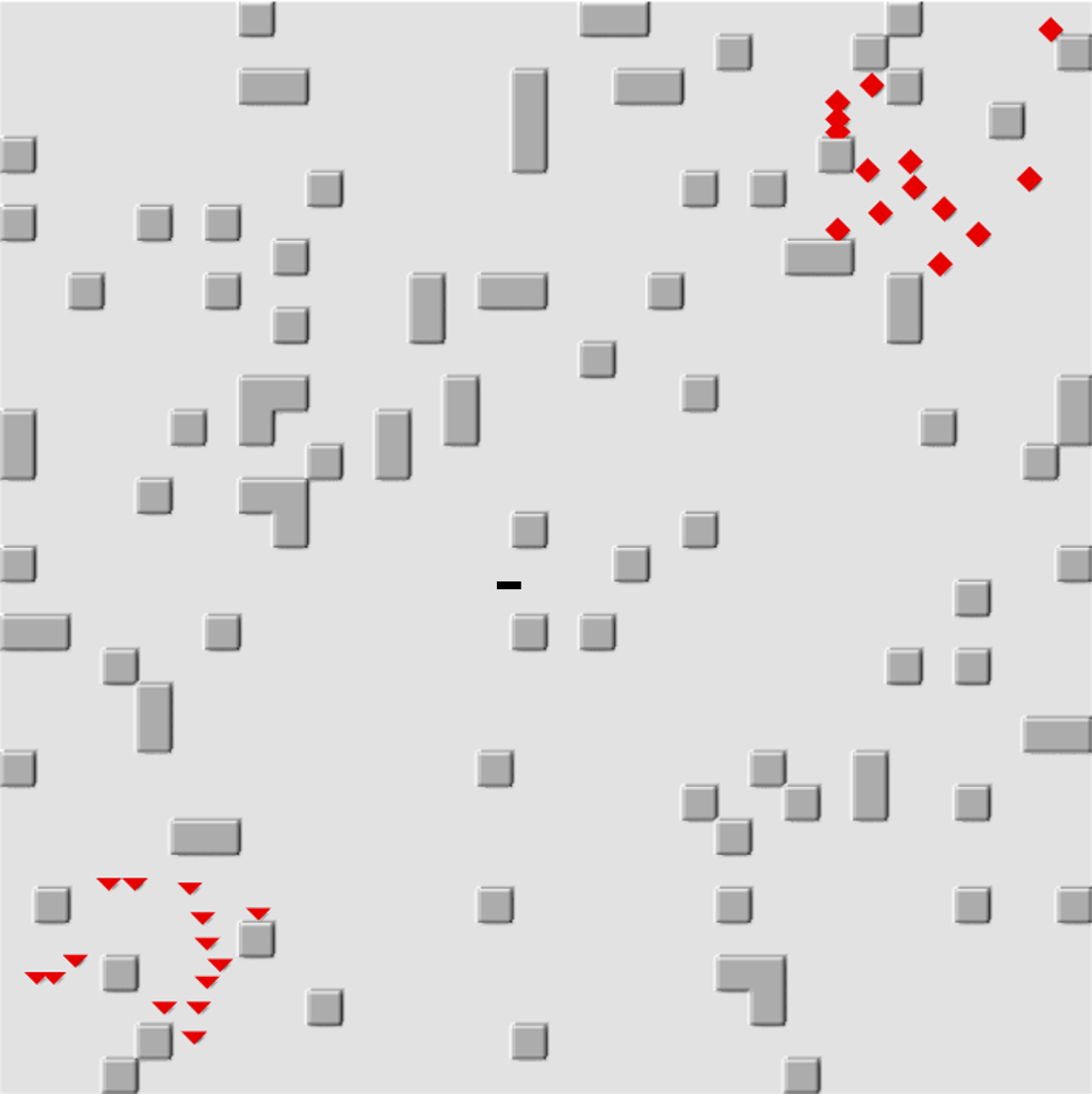}
   \label{subfig1510}
 }
 \subfigure[15 routes - Push-and-Spin]{
   \includegraphics[width=5cm] {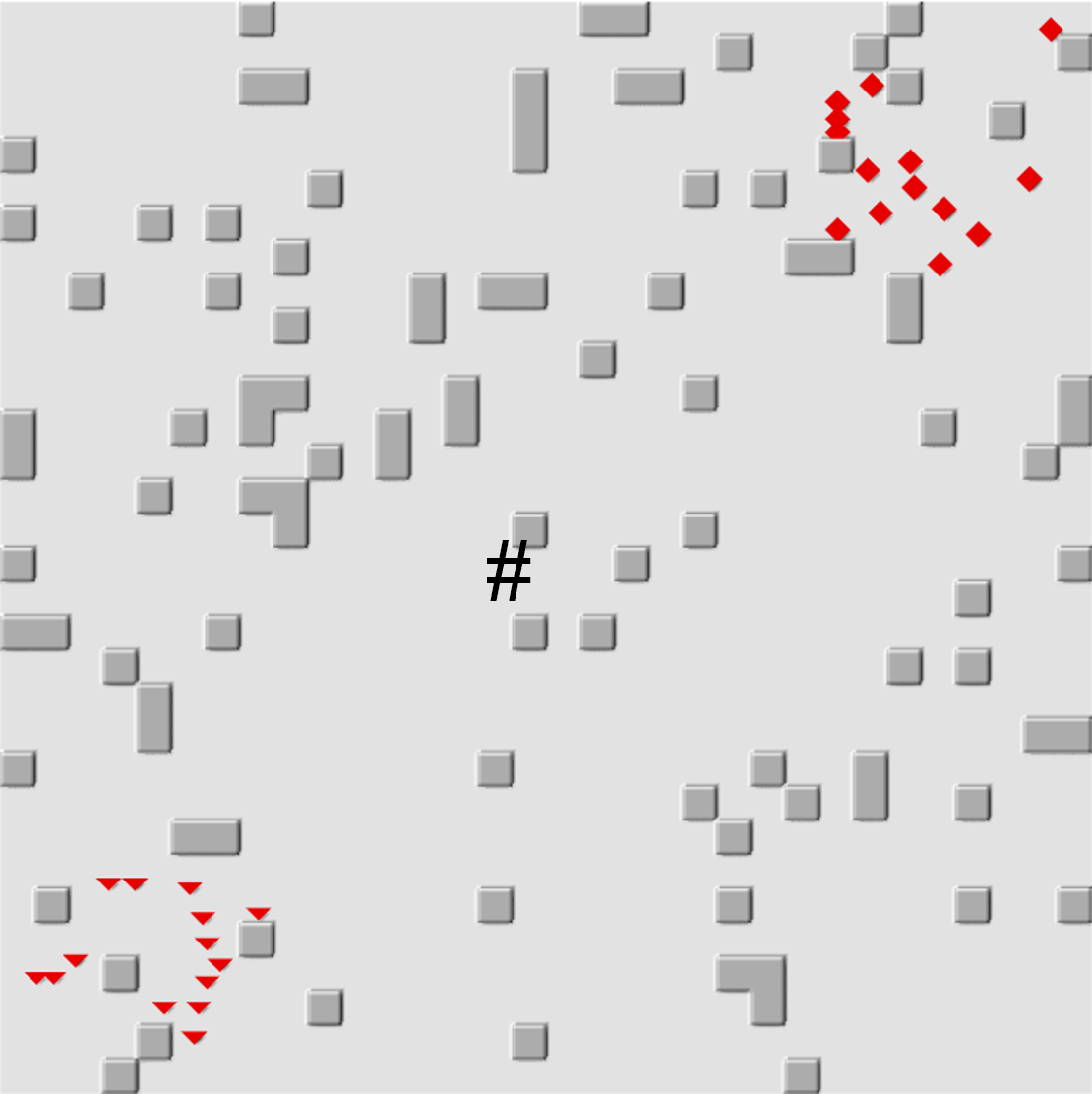}
   \label{subfig1511}
 }
 \subfigure[15 routes - The proposed method]{
   \includegraphics[width=5cm] {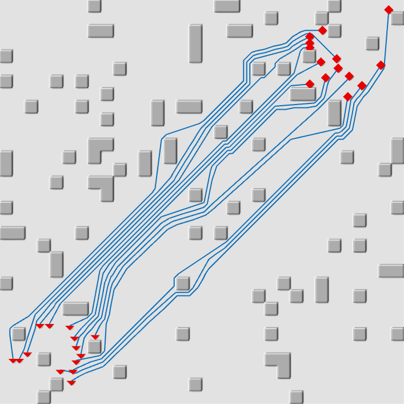}
   \label{subfig1512}
 }

  \subfigure[20 routes - CBS]{
   \includegraphics[width=5cm] {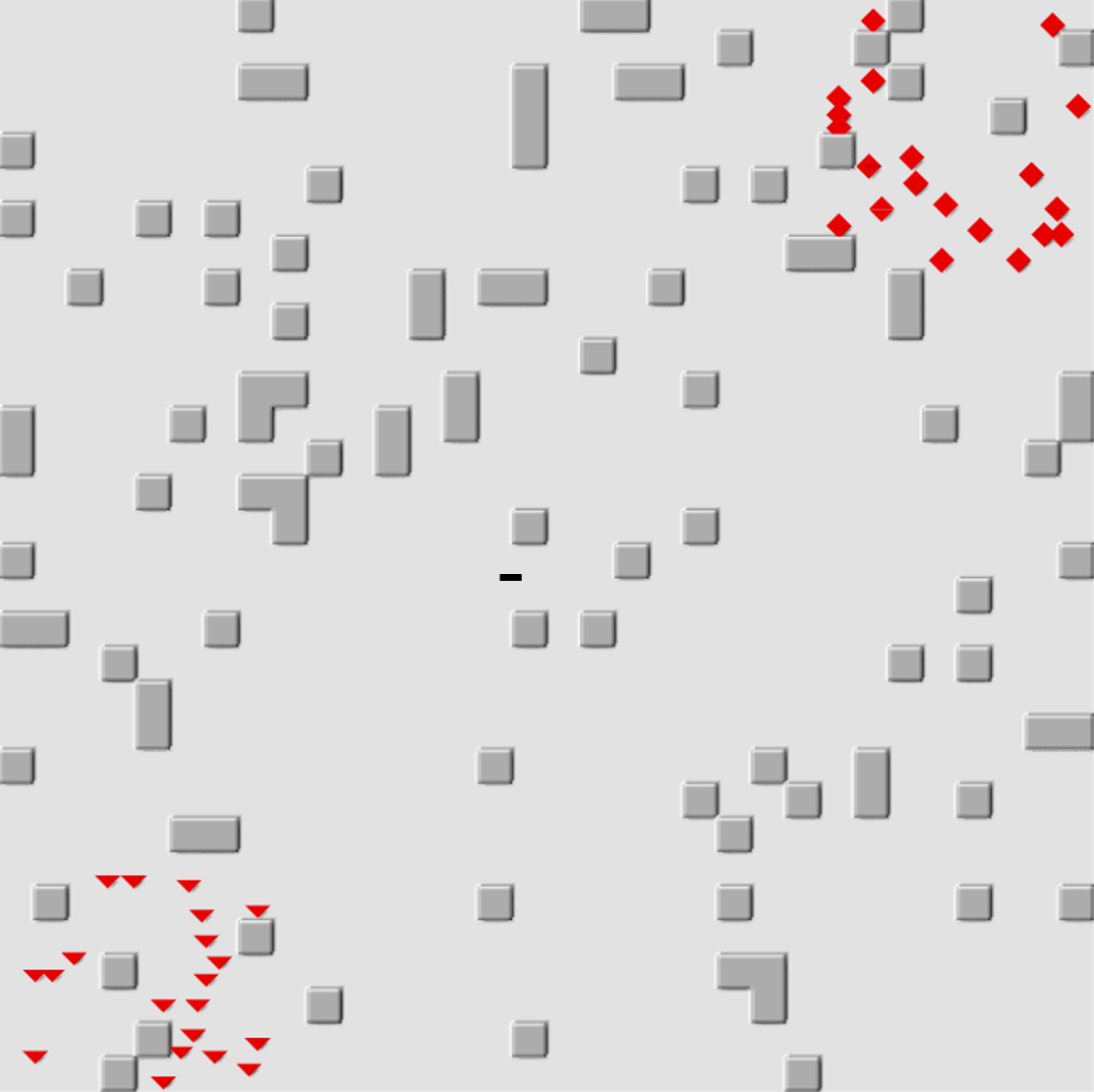}
   \label{subfig1513}
 }
 \subfigure[20 routes - Push-and-Spin]{
   \includegraphics[width=5cm] {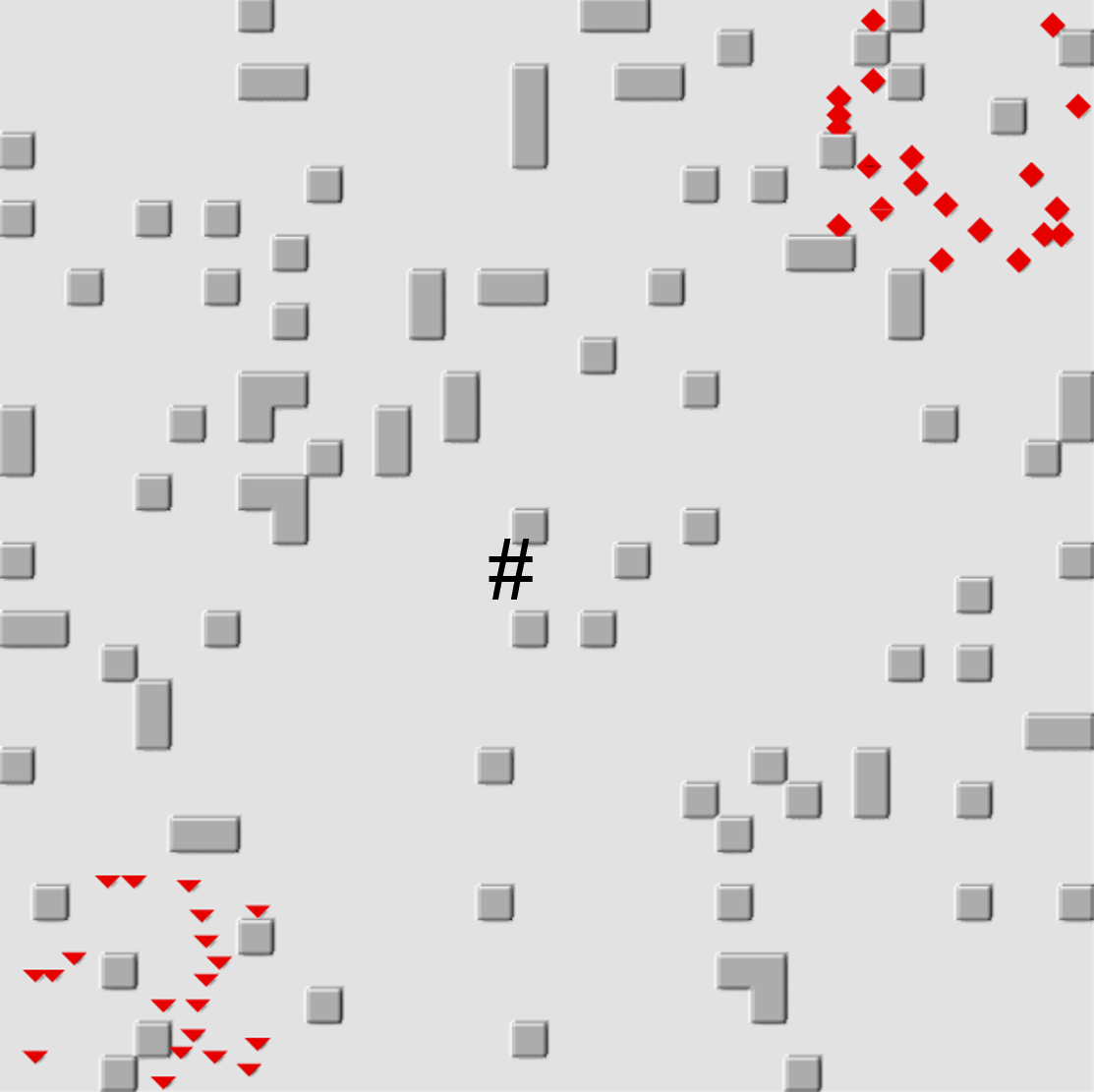}
   \label{subfig1514}
 }
 \subfigure[20 routes - The proposed method]{
   \includegraphics[width=5cm] {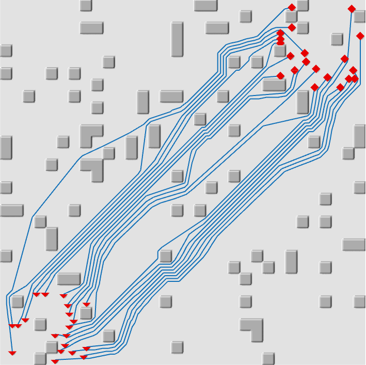}
   \label{subfig1515}
 }
 \caption{Routes planned by CBS, Push-and-Spin, and the proposed method in the 2D standard test scenario}
\label{fig15}
\end{figure}
\FloatBarrier

\subsection{Demonstration on a Real-World Scenario}
In this section, we use a real-world scenario to demonstrate the capabilty of the proposed method. The selected environment is a typical urban area in Hangzhou, China. Hangzhou has 10,711,198 residents \citep{CITYPOPULATION2020} with a size of $8,292.31km^2$ for urban districts. Hangzhou has logged more than 3 billion express parcels \citep{Hangzhou2020} in 2020. Drone delivery services have been offered by Antwork Technology in Hangzhou since 2020. The size of the selected area is $5.35km\times 2.95km$ (about $15.75 km^2$) and the details of the scenario data are shown in Table \ref{table6}. Drones are allowed to fly in the altitude range $[60m,120m]$. A graph is extracted from the scenario data using grid size $(10m,10m,10m)$. The details of the graph are shown in Table \ref{table7}. In the flyable altitude range, there is a total of $946950$ cells and $707128$ cells are traversable.

\begin{figure}[ht]
\centering
  \includegraphics[width=15cm]{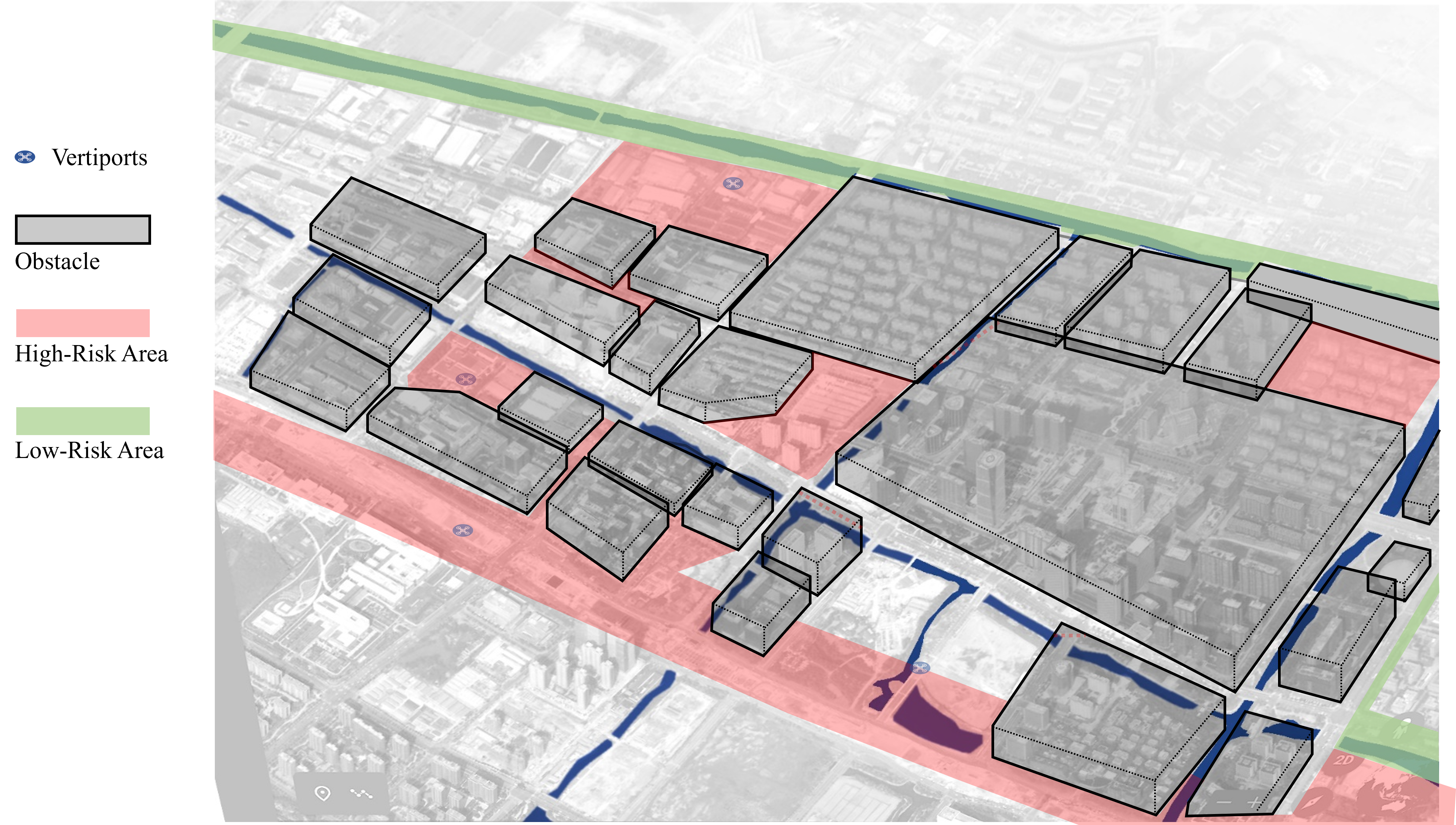}\\
  \caption{A real-world scenario in Hangzhou, China}
\label{fig16}
\end{figure}
\FloatBarrier

\begin{figure}[ht]
\centering
 
 \subfigure[Perspective view of all planned paths]{
   \includegraphics[width=13.5cm] {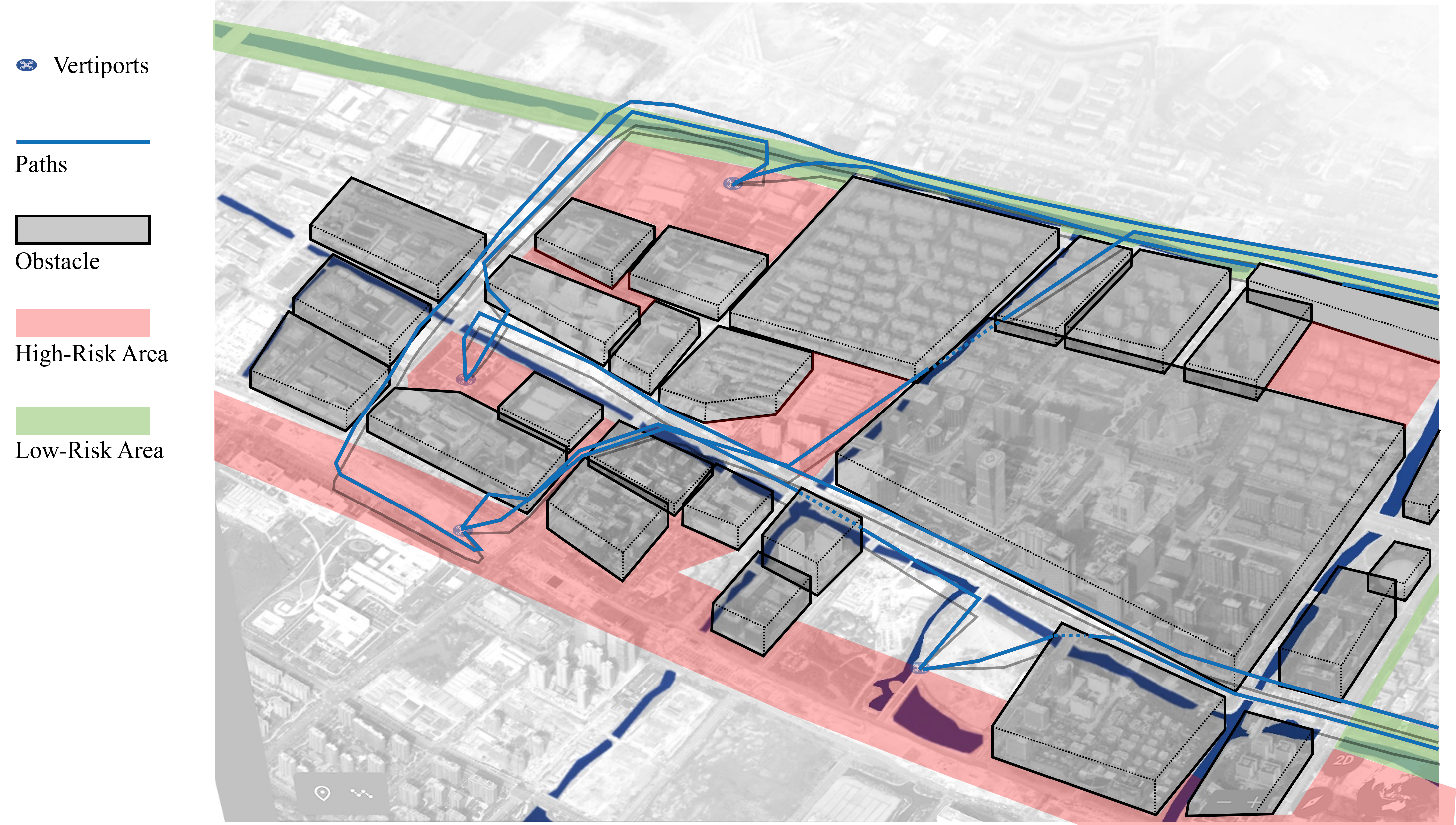}
   \label{subfig173}
 }
  \subfigure[A zoom-in view of the planned paths in relation to obstacles, high risk areas, and low risk areas]{
   \includegraphics[width=13.5cm] {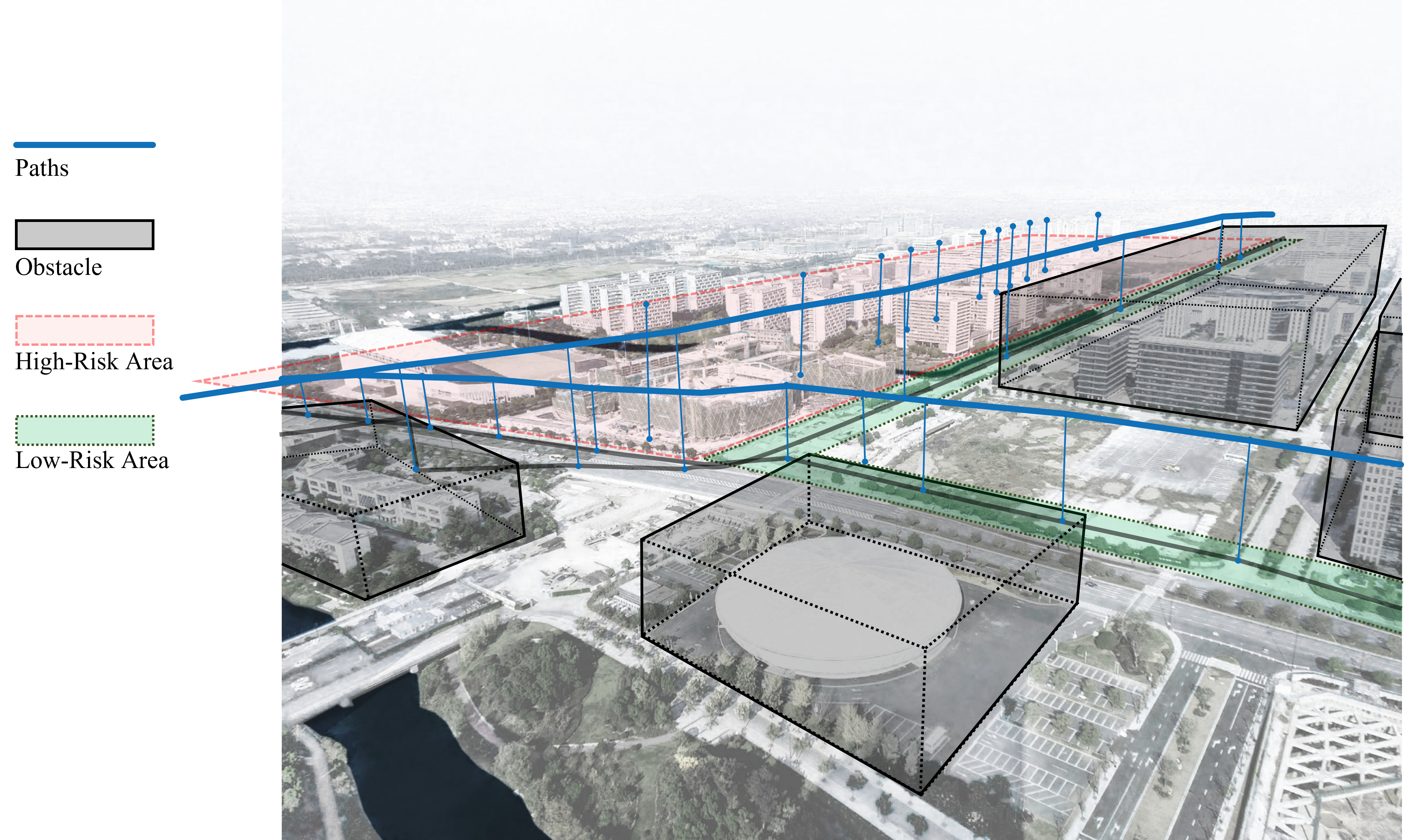}
   \label{subfig172}
 }
 \caption{A set of paths planned by the proposed method (3D)}
 \label{fig17}
\end{figure}

In this urban area, 12 routes are planned using our proposed algorithm with the following parameter settings: $\varepsilon_v=1000,\omega_r=\omega_p=1.0$. The planned route network is shown in Figure \ref{fig17}, where paths are colored in blue. A few routes are aligned together with overlapped buffer zones, showing improved airspace utilization. Many paths are over green areas, which are preferred areas for drone operations with low risk. Only a few paths are over red areas, which have high risk, as the paths over normal regions are blocked by buildings. 
With an separate validity check, we confirm that the route network generated by the proposed method satisfies all operational constraints and can be used for drone delivery services.

\FloatBarrier

\subsection{Sensitivity Analysis of Algorithm Parameters}
\subsubsection{Sensitivity Analysis on Space Cost}

In this section, we analyze the effectiveness of adding the space cost into the algorithm by testing different values of the space cost weight coefficient $\omega_p$ in solving the network planning problem in the real-world 3D scenario in Section 5.3.

We first compare the routes found without space cost and the ones found with space cost. The results are shown in Table \ref{table3}. Compared to routes found without space cost, the number of buffer zone cells decreased from 17.26 \textperthousand \ to 13.87 \textperthousand, and the number of total occupied cells reduced from 24.14 \textperthousand \ to 21.51 \textperthousand. Therefore, the proposed method with space cost can generate a route network that utilizes airspace more effectively than without space cost.

\begin{table}[ht]
    \centering
    \begin{threeparttable}
    \caption{Airspace occupancy for routes found without space cost vs. with space cost \tnote{\#}}
    \label{table3}
    
    \begin{tabular}{c| c c c}
        \hline
          & buffer zone cells & Path cells & Total occupied cells \\
        \hline
        Without space cost ($\omega_p=0$) & 17.26 \textperthousand & 6.88 \textperthousand & 24.14 \textperthousand \\
        Add space cost item ($\omega_p=1.0$)     & 13.87 \textperthousand & 6.89 \textperthousand & 21.51 \textperthousand \\
        \hline
    \end{tabular}
    \begin{tablenotes}
    	\footnotesize
    	\item[\#] results are presented as the number of cells divides the total number of cells available for drones (707128)
    \end{tablenotes}
   \end{threeparttable}
\end{table}
\FloatBarrier

\begin{figure}[ht]
\centering
\subfigure[Total occupied airspace (buffer zone vs. path) ]{
   \includegraphics[width=7cm] {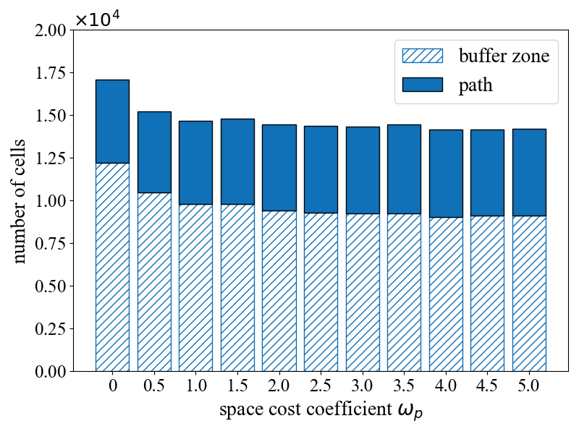}
   \label{subfig181}
 }
 \subfigure[Total occupied airspace vs. total risks]{
   \includegraphics[width=7cm] {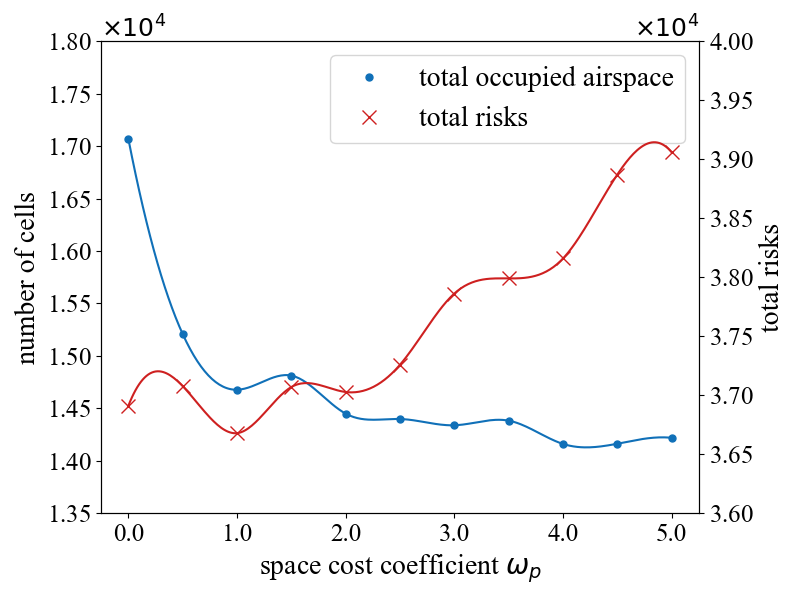}
   \label{subfig182}
 }
 \caption{Sensitivity analysis on space cost coefficient $\omega_p$}
\label{fig18}
\end{figure}

Then we test how $\omega_p$ affects the planning of a route network. The results are shown in Figure \ref{fig18}. As $\omega_p$ increases, the number of buffer zone cells and the number of total occupied cells decrease; meanwhile, the number of path cells increases slightly, and the total cost of risk increases gradually as the relative weight on risks reduces compared to weight on airspace occupancy, resulting some routes fly over high risk areas to reduce space cost. Therefore, the relative value of the space cost coefficient in relation to other cost coefficients should be carefully calibrated so that the safety aspect is not compromised. 

\FloatBarrier

\subsubsection{Sensitivity Analysis on Route Prioritization Threshold}
In this section, we first test the route network planning method with and without Route Prioritization module. Then we analyze how different values of the Route Prioritization threshold affect the results.

The result of comparison between no route prioritization and with route prioritization is shown in Table \ref{table4}. With route prioritization, routes in Urgent and Important levels have smaller costs, but routes in Normal and Low levels have larger costs. The increase in normal and low levels is expected because route prioritization improves the quality of high-priority routes at the cost of low-priority routes. There is minimal impact on the total cost of the entire network from using route prioritization.

\begin{table}[ht]
    \centering
    \caption{Comparison between no route prioritization and with route prioritization}
    \label{table4}
    \begin{threeparttable}
    \begin{tabular}{c| c c c c | c}
        \hline
          Average total cost after running 10 times\tnote{\#} & Urgent & Important & Normal & Low & Total\\
        \hline
        No Route Prioritization & 16030 & 4270 & 5572 & 11429 & 37301\\
        With Route Prioritization\tnote{*} & 15198 & 4091 & 6091 & 12331 & 37711 \\
        \hline
        Improvement after route prioritization & 832 & 179 & -519 & -901 & -410 \\
        \hline
    \end{tabular}
    \begin{tablenotes}
    	\footnotesize
    	\item[\#] Sequences without prioritization vary, the average cost is used.
    	\item[*] Thresholds $\varepsilon_v=1000$, the range of profit $v$ is $[-219, 9765]$. 
    \end{tablenotes}
   \end{threeparttable}
\end{table}
\FloatBarrier

For the sensitivity analysis on threshold values, we take $\varepsilon_v$ as an example to show how threshold values affect the route prioritization and the final result. Here we use 16 OD pairs, and the potential profit values of these OD pairs are derived from a normal distribution $v \sim \mathcal{N}(5000,2000)$. We test the threshold values $\varepsilon_v\in\{100,400,800\}$. As shown by the results in Table \ref{table5}, with the increase of $\varepsilon_v$, OD pairs are divided into fewer but larger subsequences, making the total number of possible ordered arrangements $S_g$ larger, and the minimum total costs for the networks decrease. This is because with more permutation of OD sequences being tested, the results solution gets closer to the optimal one. However, it is at the cost of increased computational time.

\begin{table}[ht]
    \centering
    \caption{Sensitivity Analysis on Threshold of Expected Profit Values $\varepsilon_v$}
    \label{table5}
    \begin{tabular}{p{1cm}<{\centering} p{1cm}<{\centering} p{1cm}<{\centering} p{1cm}<{\centering} p{1cm}<{\centering} p{1cm}<{\centering} p{1cm}<{\centering} p{1cm}<{\centering} p{1cm}<{\centering}}
        \hline
        \multicolumn{9}{l}{\textbf{OD pairs sorted by expected profit values $v$}} \\
        route $i$ & $r_1$ & $r_2$ & $r_3$ & $r_4$ & $r_5$ & $r_6$ & $r_7$ & $r_8$ \\
        $v$ & $9481$ & $8735$ & $7988$ & $7908$ & $6957$ & $6900$ & $6522$ & $5821$ \\
        route $i$ & $r_9$ & $r_{10}$ & $r_{11}$ & $r_{12}$ & $r_{13}$ & $r_{14}$ & $r_{15}$ & $r_{16}$ \\
        $v$ & $5800$ & $5667$ & $5626$ & $5423$ & $4793$ & $4697$ & $3045$ & $-105$ \\
        \hline
        \multicolumn{9}{l}{$\mathbf{\varepsilon_v=100}$} \\
        \multicolumn{3}{l}{Best total costs: 35098} & \multicolumn{3}{l}{Shuffled times: $K=10$} & \multicolumn{3}{l}{Total randomness: $S_g=32$} \\
        \multicolumn{9}{l}{subsequences: $[(r_1),(r_2),(r_3,r_4),(r_5,r_6),(r_7),(r_8,r_9),(r_{10},r_{11}),(r_{12}),(r_{13},r_{14}),(r_{15}),(r_{16})]$} \\
        \hline
        \multicolumn{9}{l}{$\mathbf{\varepsilon_v=400}$} \\
        \multicolumn{3}{l}{Best total costs: 33810} & \multicolumn{3}{l}{Shuffled times: $K=100$} & \multicolumn{3}{l}{Total randomness: $S_g=960$} \\
        \multicolumn{9}{l}{subsequences: $[(r_1),(r_2),(r_3,r_4),(r_5,r_6),(r_7),(r_8,r_9,r_{10},r_{11},r_{12}),(r_{13},r_{14}),(r_{15}),(r_{16})]$} \\
        \hline
        \multicolumn{9}{l}{$\mathbf{\varepsilon_v=800}$} \\
        \multicolumn{3}{l}{Best total costs: 33659} & \multicolumn{3}{l}{Shuffled times: $K=200$} & \multicolumn{3}{l}{Total randomness: $S_g=1440$} \\
        \multicolumn{9}{l}{subsequences: $[(r_1,r_2),(r_3,r_4),(r_5,r_6,r_7),(r_8,r_9,r_{10},r_{11},r_{12}),(r_{13},r_{14}),(r_{15}),(r_{16})]$} \\
        \hline
    \end{tabular}
\end{table}
\FloatBarrier

\subsection{Algorithm Scalability and Computational Time}
In this section, we first show how the computational time of the algorithm is affected by the random shuffle times $K$, and then we show how the computational time of the algorithm increases as the number of routes increases. The test scenario is the urban area in Hangzhou as shown in Section 5.3. All experiments are performed on a platform Intel(R) Xeon(R) Gold 5218 CPU @ 2.30GHz.  

\begin{figure}[ht]
\centering
  \includegraphics[width=8cm]{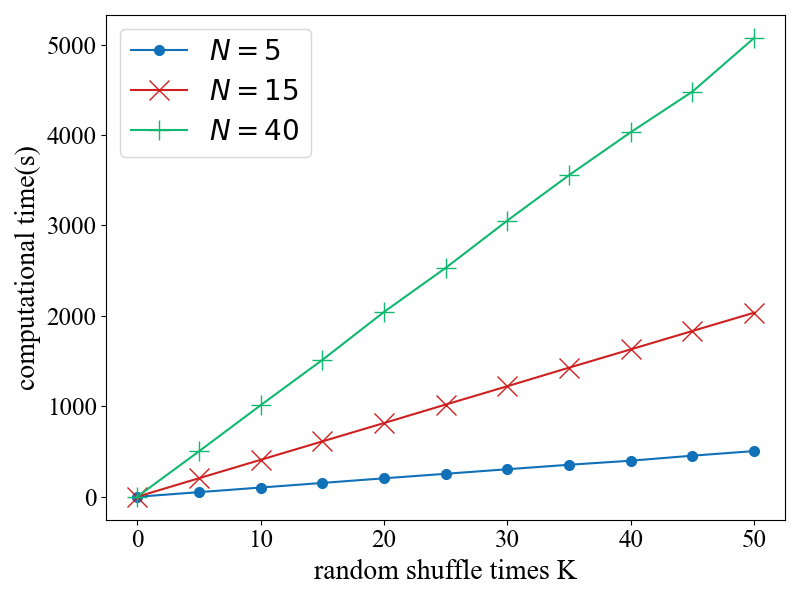}\\
  \caption{Sensitivity analysis on random shuffle times $K$ for different number of routes $N$}
\label{fig19}
\end{figure}

The computational times for different $K$ are shown in Figure \ref{fig19}. As $K$ increases, the computational time increases proportionally. This means that the total computational time is proportional to the number of OD pair sequences to run. The computational times for different number of routes to plan are shown in Figure \ref{fig20} and the result data are summarized in table \ref{table8} in the Appendix. When $K$ is fixed, with the increase of the number of routes $N$, the computational time increases near linearly. The empirical results show that the proposed method is able to handle the planning of 40 routes within about 1 hour for a real-world scenario.

\begin{figure}[ht]
\centering
  \includegraphics[width=8cm]{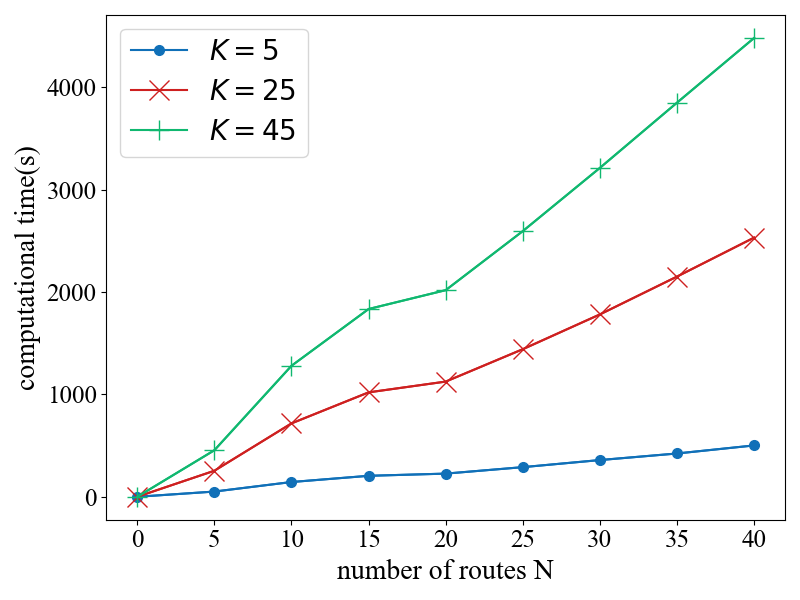}\\
  \caption{Algorithm scalability as the number of routes increases}
\label{fig20}
\end{figure}
\FloatBarrier

Antwork Technology's current drone delivery network operates up to 80-100 flights per day on 50 unidirectional air routes. The length of the air routes is about 10 km on average, and the longest one is more than 30 km. These air routes are manually designed at Antwork Technology. To design a route, a field study needs to be conducted first to obtain 3D modelling of the environment. Based on the detailed 3D model, an air route is manually charted and then checked to see whether it satisfies all operational requirements. It usually takes 2 to 4 hours to design one air route. Moreover, it becomes infeasible to design a network with a large number of air routes once the complexity exceeds human operators' capability. 

The proposed method is expected to significantly improve the design process of the air route network by making it automatic. With the proposed method, the design of a network with 40 air routes within 1 hour. Also, the scale of the network will no longer be limited by human operators' capability of dealing with the computational complexity. Using the proposed method, more routes can be designed until the airspace cannot accommodate any additional air routes.

\section{Conclusion}
This paper proposed a sequential route network planning method to support tube-based operations of drone deliveries. The proposed method is capable of designing a spatially separated tube-based route network in a large urban area within a reasonable time. The method is composed of four modules: Environment Discretization, Individual Route Planning, Route Prioritization, and Network Planning. The proposed prioritization structure decoupled the network planning problem into a sequential single-path-planning problem. The space cost function in the Individual Route Planning module made it possible to have the routes aligned and grouped together. A set of tests were carried out to show our method can generate route networks that satisfy all requirements within a short time and can be taken for commercial use. With the implementation of the proposed method, drone delivery service providers can quickly design a drone route network, and re-plan on a daily basis to respond to changes in their service network. With the route prioritization function, they can prioritize the design of urgent or important deliveries. The space cost function allowed higher airspace utilization and potentially led to the identification of high-volume air corridors in an urban area. From the city residents' perspective, less areas would be affected by this new type of aerial operations. 

One direction of future work is to further reduce computational time while ensuring the optimality of the solution and the fairness between different routes. Distributed planning is a promising direction to explore.

Another direction to adapt the method to build tube-based route networks for other applications, where many moving agents should be transported among OD pairs. In the tube-based route network, one or multiple tubes can be planned to connect an OD pair based on the traveling demand. The moving agents following a tube can sequentially pass to arrive at the destination. Besides the drone route network, an example situation is warehouse logistics. In such a situation, robots should move packages from one place to another place, and the proposed method can be applied to plan the paths for these robots.

\bibliographystyle{model5-names}
\bibliography{ref}

\section*{Appendix A: Calculation for $\lambda_r$ and $\lambda_p$}
$\lambda_r$ is calculated as $\lambda_r=\frac{o_0(s_{start},s_{goal})} {r_0(s_{start},s_{goal})}$, where $o_0(s_{start},s_{goal})$ is the minimum operational cost and $r_0(s_{start},s_{goal})$ is the minimum risk cost for a conflict-free path that connects $s_{start}$ and $ s_{goal}$.

$\lambda_p$ is calculated as $\lambda_p=\frac{o_0(s_{start},s_{goal})} {p_0(s_{start},s_{goal})}$, where $o_0(s_{start},s_{goal})$ is the minimum operational cost and $p_0(s_{start},s_{goal})$ is the minimum space cost for a conflict-free path that connects $s_{start}$ and $ s_{goal}$.

\section*{Appendix B: Real-World Scenario Data Specifications}

\begin{table}[ht]
    \centering
    \caption{Description for real-world scenario data}
    \label{table6}
    \begin{tabular}{l l l}
        \hline
        Types & Number & Attributes \\
        \hline
        Vertiports & 7 & \textit{Id, type, radius, \{latitude, longitude, altitude\}} \\ 
        Obstacles & 55 & \textit{Id, highest / lowest altitude, points (latitudes, longitudes)} \\
        High / low risk area & 22 & \textit{Id, risk level, points (latitudes, longitudes)} \\
        \hline
    \end{tabular}
\end{table}
\FloatBarrier

\begin{table}[ht]
    \centering
    \caption{Summary for real-world scenario}
    \label{table7}
    \begin{tabular}{l l l}
        \hline
        Name & Value & Example \\
        \hline
        \multirow{3}{0.3\textwidth}{Bounding box} & X-direction & $[-1780.99m,3566.63m]$\\
        & Y-direction & $[-1177.21m,1768.74m]$\\
        & Z-direction & $[0m,260m]$\\
        \hline
        Traversable layer & Z-direction & $[60m,120m]$ \\
        \hline
        \multirow{4}{0.3\textwidth}{Map size} &Total area & $15.75 km^2$\\
        & Grid size & $(10m,10m,10m)$\\
        & Number of grids & $535*295*6=946950$\\
        & Number of traversable grids & $707128 (74.7\%)$\\
        \hline
        \multirow{6}{0.3\textwidth}{Cell attributes} & \multirow{3}{0.3\textwidth}{Risk} & $(0,1)\to$ \textit{low risk area}\\
        & & $(1,\infty)\to$ \textit{high risk area} \\
        & & $1 \to$ \textit{normal area} \\
        & Reachable & \textit{True} or \textit{False} \\
        & Reserved & \textit{True} or \textit{False} \\
        & Buffer & \textit{True} or \textit{False} \\
        \hline
    \end{tabular}
\end{table}
\FloatBarrier

\section*{Appendix C: Computational time for different Random shuffle Times $K$ and number of routes $N$}

\begin{table}[ht]
    \centering
    \caption{Computational time (s) for different Random shuffle Times $K$ and number of routes $N$}
    \label{table8}
    \begin{tabular}{|c | c c c c c c c c c c c|}
    \hline
    \diagbox{N}{K} & 0 & 5 & 10 & 15 & 20 & 25 & 30 & 35 & 40 & 45 & 50\\
    \hline
    0 & 0 & 0 & 0 & 0 & 0 & 0 & 0 & 0 & 0 & 0 & 0\\
    5 & 0 & 51.0 & 102.0 & 153.1 & 204.2 & 254.3 & 304.6 & 354.1 & 399.0 & 454.6 & 505.5 \\
    10 & 0 & 144.1 & 286.3 & 426.5 & 579.5 & 716.7 & 858.1 & 996.3 & 1127.1 & 1279.3 & 1430.3 \\
    15 & 0 & 204.8 & 408.5 & 612.2 & 815.2 & 1019.7 & 1222.6 & 1427.8 & 1630.3 & 1833.6 & 2035.6 \\
    20 & 0 & 226.2 & 450.7 & 673.7 & 898.3 & 1124.9 & 1347.7 & 1570.4 & 1795.0 & 2017.8 & 2242.8  \\
    25 & 0 & 289.3 & 577.2 & 864.0 & 1152.6 & 1441.3 & 1729.8 & 2018.4 & 2307.0 & 2594.5 & 2885.1 \\
    30 & 0 & 358.6 & 712.0 & 1067.0 & 1426.0 & 1781.0 & 2139.2 & 2498.0 & 2854.6 & 3212.1 & 3567.2 \\
    35 & 0 & 422.8 & 847.6 & 1298.6 & 1726.8 & 2150.5 & 2578.8 & 3001.5 & 3426.0 & 3849.0 & 4272.7 \\
    40 & 0 & 501.5 & 1011.9 & 1512.5 & 2043.4 & 2531.1 & 3054.1 & 3556.6 & 4031.7 & 4480.7 & 5074.0 \\
    \hline
    \end{tabular}
\end{table}
\FloatBarrier

\end{document}